\documentclass[12pt]{article} 

\usepackage{xr}
\usepackage[margin=1in]{geometry}
\usepackage[utf8]{inputenc}
\usepackage{graphicx}
\usepackage{authblk}
\usepackage{booktabs}

\usepackage{xcite}
\usepackage{xr}

\usepackage{booktabs}
\usepackage{tabularx}
\usepackage[figuresright]{rotating}

\makeatletter
\newcommand*{\addFileDependency}[1]{
  \typeout{(#1)}
  \@addtofilelist{#1}
  \IfFileExists{#1}{}{\typeout{No file #1.}}
}
\makeatother

\newcommand*{\myexternaldocument}[1]{%
    \externaldocument{#1}%
    \addFileDependency{#1.tex}%
    \addFileDependency{#1.aux}%
} 
\myexternaldocument{supp}

\usepackage{amsmath}
\usepackage{amssymb}
\usepackage{newtxmath}
\DeclareMathAlphabet{\mathpzc}{T1}{pzc}{m}{it}

\usepackage{bm}

\DeclareMathAlphabet{\pazocal}{OMS}{zplm}{m}{n}
\newcommand{\unif}{\pazocal{U}}

\usepackage{caption}
\usepackage[table, x11names, dvipsnames]{xcolor}

\definecolor{dkgreen}{rgb}{0,0.6,0}
\definecolor{gray}{rgb}{0.5,0.5,0.5}
\definecolor{mauve}{rgb}{0.58,0,0.82}

\usepackage{xurl}

\usepackage{changepage}

\usepackage{setspace}
\usepackage[font=small,labelfont=bf]{caption}

\title{What makes Individual \textit{I}'s a Collective \textit{We}; \\ Coordination mechanisms \& costs}

\date{\today}

\author[1,2,3 ]{Jisung Yoon}
\author[4]{Vicky Chuqiao Yang}
\author[5]{Chris Kempes}
\author[2, 3]{Seoul Lee}
\author[5]{Geoffrey West}
\author[2,4,5,*]{Hyejin Youn}

\affil[1]{KDI School of Public Policy and Management, Sejong, Republic of Korea}
\affil[2]{Kellogg School of Management, Northwestern University, Evanston, IL, USA}
\affil[3]{Northwestern Institute on Complex Systems, Evanston, IL}
\affil[4]{MIT Sloan School of Management, Massachusetts Institute of Technology, Cambridge, MA, USA}
\affil[5]{Santa Fe Institute, Santa Fe, NM, USA}
\affil[*]{Correspondence can be sent to hyejin.youn@kellogg.northwestern.edu.}

\date{\today}

\begin{document}

\maketitle

\begin{abstract}
The collective effort exceeds the sum of its parts when individuals coordinate and regulate their activities and behaviors. This holds true even in self-organizing systems with open, voluntary participation where coordination occurs implicitly. Here, we analyze the non-functional actions of contributors, administrators, and bots on Wikipedia, categorizing them by their asymmetric authority: one-way oversight and two-way.  This categorization helps us reveal comparable patterns. First, we find remarkably consistent scaling factors for each category relative to system size. Two-way coordination scales superlinearly (with an exponent of $1.3$), while oversight coordination grows sublinearly (with an exponent of $0.9$), suggesting an underlying mechanism for coordination across communities. Second, we identify the hierarchical modular structure of interactions as a key factor for the economy of scale in coordination, and we propose a mathematical model to explain these results. Finally, our temporal analysis shows a shift from two-way interactions to one-way oversight as system size increases. This suggests the emergence of a nascent hierarchical structure even in self-organizing systems, echoing Weber's theory of organizational evolution.

\end{abstract}

\baselineskip24pt

\section*{Introduction}

In complex systems, ranging from biological organisms to human organizations, coordination mechanisms are crucial for understanding how numerous individual components come together to form a cohesive and operational whole across various scales~\cite{malone2018superminds, tuomela2007philosophy, weinberg1994}. What mechanisms transform isolated individual \textit{I}s into a cohesive collective \textit{We}? Consider a bustling bee colony, a remarkable super-organism that operates seamlessly. Beyond their genetic predispositions, these insect societies require constant coordination to adapt to complex environments~\cite{huang1992honeybee}. This includes frequent role adjustments~\cite{moffett2021ant}, communication through pheromones~\cite{slessor2005pheromone}, dances~\cite{dyer2002biology}, and vibrations~\cite{schneider2004vibration}, as well as resolving occasional conflicts~\cite{galbraith2016testing}. Human society is not too different, facing similar coordination challenges. In the company, over the past two decades, the time devoted to coordination by managers and employees has increased by 50\% or more~\cite{collabhbr}. In academia, a recent survey indicates that university faculty spend about 45\% of their weekly work hours on coordination, such as meetings, emails, scheduling, planning, and administrative tasks, which are not directly related to the core functions of academia~\cite{Ziker2014bluereview}.
As our society becomes more complex, necessitating larger teams with experts in specialized skills, this trend is expected to continue growing~\cite{wuchty2007increasing,hosseinioun2023deconstructing}. Yet, in complex systems, coordination processes and their associated cost often do not immediately appear~\cite{toxtli2021quantifying}, and there are fundamental questions yet to be answered about how much of this coordination function is needed, and what underlying mechanisms generate the coordination cost we observe.

What are the determining factors behind the cost of coordination mechanisms within a collective? Several elements come into play, including the size of the organization~\cite{camacho1991adaptation, zhang2011group}, its structure~\cite{marsden1994organizational, powell2003neither}, and the complexity of the problems it grapples with~\cite{thompson2003organizations, straub2023cost, zhou2013designing}. Moreover, these factors often co-evolve. For instance, economies of scale are frequently considered in coordination costs. As an organization grows, the relative proportion of administrators may decrease, potentially reducing costs~\cite{klatzky1970relationship, blau1970formal}. Simultaneously, with organizational growth, organizations tend to adopt impersonal coordination mechanisms such as formalization, standardization, and hierarchical control to enhance efficiency~\cite{klatzky1970relationship}.
However, the co-evolution of various factors and the intertwined types of coordination that arise as more individuals (\textit{I}) join the collective (\textit{We}) remain largely unexplored~\cite{dedeo2014group}.




To empirically assess the coordination costs and unpack its underlying mechanism, we chose Wikipedia as our focal point. 
Wikipedia is an ideal system for studying these dynamics because of its vast collection of individual activity records and myriad small communities associated with individual projects, continually evolving through contributions from a diverse array of intellectual minds~\cite {yasseri2012dynamics, yun2019early}. Individual projects are defined by groups of varying sizes, giving us a perfect window into various scales. 
Yet, each project shares the task of constructing a comprehensive knowledge structure on a specific topic. Within this community, individuals contribute their knowledge collaboratively to build a cohesive knowledge structure for each project~\cite{zhu2013effects, klapper2023peer}. This process requires both \textit{symmetric} coordination, such as communication on talk pages~\cite{kittur2008harnessing, kittur2010beyond, aaltonen2015building} and decisions about others' edits~\cite{yasseri2012dynamics, tsvetkova2016dynamics, zhang2019participation, halfaker2011don, dedeo2016conflict}, and \textit{asymmetric} coordination, such as actions by authorities~\cite{arazy2019neither, danescu2012echoes, greenstein2021ideology}, the emergence of norms~\cite{heaberlin2016evolution}, and even the enforcement of norms by automated bots~\cite{geiger2009social, steiner2014bots, tsvetkova2017even}. Notably, Wikipedia is entirely authored and maintained by a decentralized community of volunteers, embodying the essence of collective intelligence. For this reason, Wikipedia is considered more like a community or ecosystem rather than a traditional organization with no formal organization structure in the classification spectrum~\cite{malone2018superminds}. However, as we will demonstrate later, Wikipedia exhibits a \textit{nascent} hierarchical structure~\cite{shaw2014laboratories, yun2019early}, which can significantly enhance the effectiveness of coordination within its operations.

In this paper, first, we investigate how coordination mechanisms evolve when the number of contributors expands within each project by analyzing how various metrics scale with the number of contributors. In almost all cases, we find robust evidence of simple power-law scaling (see Methods) whose exponents quantitatively reveal that the nature of coordination mechanisms systematically shifts as the number of contributors on a project increases. Furthermore, based on these scaling exponents, we are able to quantify comparisons between our results and other systems, including biological~\cite{koonin2006power, kempes2017drivers}, ecological ~\cite{cody1975ecology}, and urban systems~\cite{bettencourt2007growth, youn2016scaling}. 
Our study reveals that symmetric (two-way) coordination exhibits a superlinear scaling relationship with the number of contributors (i.e., interactions per contributor \textit{increases}  as the size of the project increases), while asymmetric (one-way) coordination increases sublinearly as contributor numbers rise (i.e., the amount of supervision per contributor \textit {decreases} as the size of the project increases, 
reflecting an economy of scale). Second, we discovered that modules play a crucial role in explaining the emergence of economies of scale in supervision and rule-based coordination. Third, we identify a trade-off between coordination mechanisms, primarily categorized into personal and impersonal coordination. With this categorization, we measure the extent to which each project depends on personal or impersonal coordination. We found that, as time progresses, projects tend to transition towards impersonal coordination, echoing Weber's theory of organizational formalism~\cite{weber2023bureaucracy}. Lastly, we propose a mathematical model of the coordination mechanism, explaining the interplay between one-way coordination and two-way coordination,  emphasizing modular structure and organizational learning.

\begin{figure*}[t]
    \centering
    \includegraphics[width=.97\textwidth]{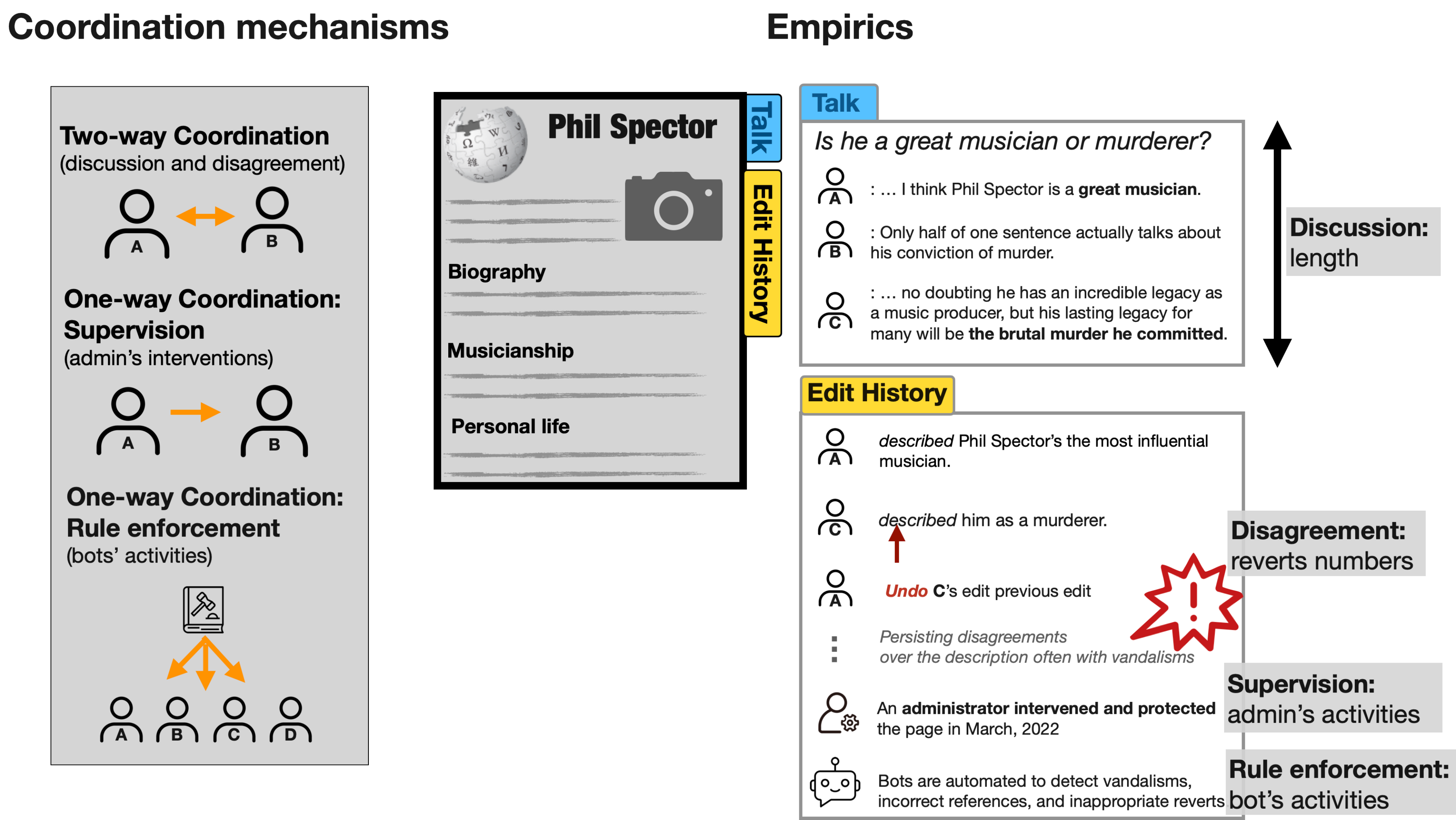}
    \caption{\textbf{Coordination Mechanisms in Wikipedia} We illustrate coordination mechanisms in Wikipedia with the Phil Spector project as an example. On the associated talk pages, contributors discuss which of Phil Spector's musical achievements and his crime should be highlighted. Within the edit history, a pattern of successive reversions by contributors in similar contexts becomes evident. Since there is no inherent hierarchy among contributors regarding discussions and reversions, conflicts can endure over extended periods and sometimes escalate into acts of vandalism. In response, human administrators intervene and protect the project by temporarily restricting editing to mitigate ongoing conflicts and restore order. Lastly, automated bots monitor and enforce established rules by detecting obvious vandalism, correcting references, and addressing inappropriate reverts.}
    \label{fig:schem}
\end{figure*}


\section*{Measuring Coordination Cost in Wikipedia}


The goal of Wikipedia is to capture the entirety of human understanding as manifested by over 6 million projects covering an expansive range of topics. To achieve this grand ambition, contributors volunteer to edit segments for each project (or article/page), aiming to maintain a coherent and consistent structure throughout. However, given the vast intellectual diversity of individuals, differing views on any given topic are inevitable and pursuing structural cohesion from various people comes with a price~\cite{mintzberg1980structure, weber2023bureaucracy}. To understand price of making \textit{I}'s into \textit{We} in a decentralized system, we evaluate two types of coordination costs in Wikipedia: \emph{two-way coordination} and \emph{one-way coordination} as illustrated in Fig \ref{fig:schem} focusing on asymmetric nature. Here, \textit{two-way coordination} is characterized by mutual interactions without inherent asymmetry between two or more parties. In contrast, \textit{one-way coordination} involves predominantly unilateral processes due to an inherent power differential or asymmetry between individuals. In the context of Wikipedia, two-way coordination manifests through discussions and discords among contributors of relatively equal standing, aiming to resolve differing viewpoints and achieve consensus on content. Conversely, one-way coordination involves processes such as administrative actions or rule enforcement, where certain individuals (e.g., administrators, bots) hold authority over others.

To empirically analyze the two-way coordination cost in Wikipedia, we focus on two key metrics: the number of reverts and the length of discussions. First, we examine the number of reverts --- a feature that allows contributors to completely undo the work of others and restore the article to a previous state --- as a proxy for disagreement within each project. Reverts is often considered as  norm violation~\cite{jan2017testing}, or indicators of conflicts~\cite{yasseri2012dynamics, tsvetkova2016dynamics, zhang2019participation,halfaker2011don} in Wikipedia. We exclude reverts by administrators or bots due to the inherent asymmetry in authority. Second, when simple reversals fail to resolve conflicts or when reversions occur repeatedly without resolution, extensive discussions among contributors become necessary to make decisions about the content and potential improvements of a project. We measure the length of ``talk page'', which serves as a dedicated space for discussing potential improvements to each project~\cite{kittur2008harnessing, kittur2010beyond}. Talk pages not only allow contributors to propose and debate article edits but also serve as forums to establish shared norms, guidelines, and document past conflict resolutions, often within a Frequently Asked Questions section (FAQ).

To quantify the one-way coordination cost in Wikipedia, we analyze two distinct types of interventions: supervision by human administrators and rule enforcement by automated bots. First, we examine the edit activity of administrators, who possess asymmetric privileges, as a proxy for the level of supervision. The most prominent type of administrator on Wikipedia is known as a ``Sysop" (system operator), comprising only 0.001\% of Wikipedia users~\cite{arazy2015functional}. Sysops hold significant authority to address vandalism, enforce community policies and guidelines, and mediate disputes (see Supporting Information). Sysops often delegate their responsibilities and administrative privileges to trusted Wikipedia users, known as delegated administrators. We consider these delegated administrators as administrators since they also have asymmetric privileges compared to ordinary contributors.

Additionally, we quantify the edit activity of automated bots as an indicator of impersonal rule enforcement within Wikipedia~\cite{geiger2009social, steiner2014bots}. Bots are programmed tools that execute repetitive and well-defined tasks essential for project maintenance. Wikipedia upholds a strict bot policy, primarily overseen by administrators. The objectives and activities of each bot are thoroughly documented on their respective pages. One example is anti-vandalism bots like ClueBot NG, which are specifically designed to detect and swiftly undo instances of vandalism. According to an empirical study, when ClueBot NG experienced a breakdown, the number of revert edits nearly doubled~\cite{geiger2013levee}, highlighting the bot's effectiveness in suppressing conflicts and maintaining the quality of Wikipedia's content. Bots can also generate high-quality articles by correcting evident deviations from established templates~\cite{zheng2019roles} and improving the verifiability of projects~\cite{petroni2023improving}. These automated processes complement the supervision by human administrators by maintaining order, enforcing policies, and ensuring adherence to community standards.

\begin{figure*}[!t]
    \centering
    \includegraphics[width=\textwidth]{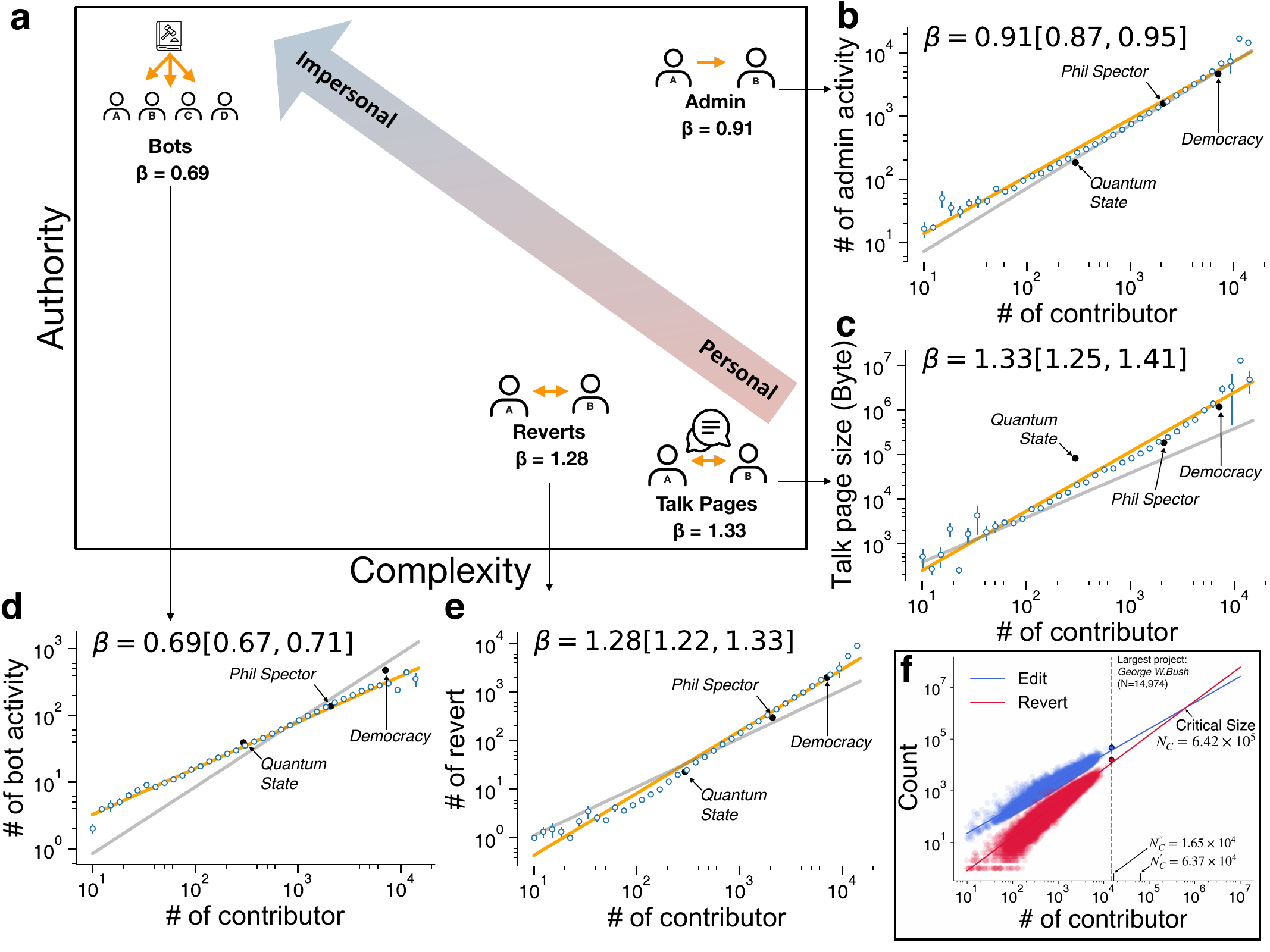}
    \caption{\textbf{Interplay of Coordination Mechanisms Across Complexity and Authority Dimensions.} \textbf{(a)} Phase diagram illustrating the regulatory functions along the authority and complexity dimensions. \textbf{(b-e)} Scaling behavior of \textbf{(b)} admin activities, \textbf{(c)} Talk page size, \textbf{(d)} automated bot actions, and \textbf{(e)} frequency of reverts with number of contributors.  Each project is log-binned (blue) with error bars and estimated $\beta$ in $N^\beta$ (the orange line), accompanied by a linear trend (gray line) for reference.  Two-way coordination such as reverts and talk page size shows superlinear scaling, growing faster than the number of contributors, while one-way coordination such as supervision by human administrator and rule enforcement by automated bots shows sublinear scaling, increasing more slowly than contributor numbers. Economy of scale transitioning from superlinear to sublinear scaling, corresponding to a shift from personal interactions in the lower right to more impersonal interactions in the upper left.  We highlight projects that we discussed in the paper: Democracy, Phil Spector, and Quantum State. \textbf{(f)} When does the Wikipedia project collapse? Scaling behavior of contributions (blue, $\beta_{\text{edit}} = 1.01$) and reversals (red, $\beta_{\text{revert}} = 1.31$) is shown here. The lines intersect at $N_c \approx 6.42 \times 10^5$, where the project becomes perfectly unproductive. Two additional crossover points, $N_c^{'} = 6.37 \times 10^4$ (reverting one in two edits) and $N_c^{''} = 1.65 \times 10^4$ (reverting one in three edits), are marked for reference, both still indicating unproductivity. The largest project in Wikipedia is marked to show how unproductive it has been.}
    \label{fig:costs_vital}
\end{figure*}

In Fig.~\ref{fig:costs_vital}a, we propose a conceptual framework to categorize coordination mechanisms along two principal dimensions: \textit{authority} and \textit{complexity}. The authority dimension distinguishes between one-way coordination with inherent power asymmetries (top) and two-way coordination involving mutual interactions among peers (bottom). The complexity dimension ranges from straightforward, easily codified tasks (left) to ambiguous, intricate tasks requiring diverse inputs (right). Aligning with this framework, reverts and discussions on talk pages frequently involve nuanced, context-dependent tasks without a clear hierarchical structure, placing them in the low authority, high complexity quadrant (bottom right). In contrast, administrators often handle complex issues while wielding authority over others, positioning administrative activities in the  high authority, high complexity quadrant (top right). Bots, however, generally execute well-defined, rule-based tasks adhering to organizational policies, situating them in the high authority, low complexity quadrant (top left).  This conceptual mapping illustrates how the nature of tasks (complexity) and the asymmetry of interactions (authority) together provide a framework for understanding the different scaling factors of each coordination cost and the interplay among coordination mechanisms, which are discussed in more detail later.

\section*{Scaling Factors in Coordination Costs}

How cost of coordination increase with organizational growth? Many assume that coordination costs naturally rise as an organization expands~\cite{camacho1991adaptation}. However, the crucial question we aim to explore is whether these costs grow at a faster or slower rate compared to the organization’s size. To address this, we adopt a scaling framework as a useful tool for quantitative assessment (See Methods) ~\cite{west2018scale}. We use an integral approach to estimate the organization size, $N$, measured as the cumulative total of all unique contributors who have ever made edits on a page (See Supporting Information). Then, we consider power-law scaling (See Methods) as follows:
\begin{equation}
    Y = Y_0N^{\beta},
    \label{eq:scaling_relation}
\end{equation}
where $Y$ represents the coordination cost,  $Y_0$ is a normalization constant. The exponent $\beta$ quantifies the rate of increase in $Y$ with relative increases in $N$ as being greater ($\beta>1$) or less ($\beta<1$) than linear.


The scaling exponent $\beta$ serves as a predictor of the system's behavior. As an illustrative example, in Wikipedia, while the number of edits increases linearly, reversals surge super-linearly as the number of contributors grow. The scaling framework predicts a critical point $N_c$, where the project faces potential `collapse' where reversals surpass edits, risking project collapse, that satisfies
\begin{equation}
    Y_{\text{edit}} (N_c) \leq Y_{\text{revert}} (N_c).
\end{equation} 
The scaling factors from the empiric are  $Y_{\text{edit}} = 2.21 \times N^{1.01}$ and $Y_{\text{revert}} = 0.04N^{1.31}$\footnote{Here, we count reverts by all contributors to the project}. Solving $2.21N_c^{1.01} = 0.04N_c^{1.31}$ gives $N_c \approx 6.42 \times 10^5$, as depicted in Fig.~\ref{fig:costs_vital}f. This is close to Washington, D.C.'s population but not immediately alarming. We also examine less extreme cases where every other edit is reverted ($N_c^{'} \simeq 6.37 \times 10^4$), and another where one is three edits is reverted ($N_c^{''} \simeq 1.65 \times 10^4$).  These critical points, around 10,000 individuals, suggest Wikipedia are nearing un-productivity, akin to the largest Wikipedia project, \textit{George W. Bush}, raising immediate concern.



Fig.~\ref{fig:costs_vital} illustrates how coordination costs scale with the number of contributors. Cost of two-way coordination, such as reverts and talk page size, exhibit a superlinear scaling relationship ($\beta_{\text{revert}}, \beta_{\text{talk}} \simeq 1.3$,  Fig.~\ref{fig:costs_vital} c,e). This implies that disagreements and discussion in Wikipedia grow faster than the number of contributors ---\emph{ descaled economy of scale}, reflecting the societal characteristics of two-way coordination driven by interactions, which may follow mechanisms similar to those driving urban features with superlinear scaling~\cite{bettencourt2007growth, youn2016scaling}. In contrast, cost of the one-way coordination, including administrative activities and bot activities increases sublinearly, $\beta_{\text{adm. act.}} \simeq 0.9$ and $\beta_{\text{bot.act.}} \simeq 0.7$, respectively (Fig.~\ref{fig:costs_vital} b, d), showing \emph{economies of scale}, consistent with economies of scale of infrastructure in urban systems~\cite{bettencourt2007growth} and administrative staff in companies~\cite{klatzky1970relationship}. The sublinear scaling relationship is robust to measuring the number of unique administrators and bots (Fig.~\ref{fig:costs_number}), $\beta_{\text{adm.}} \simeq 0.7$ and $\beta_{\text{bot}} \simeq 0.4$, respectively, supporting natural development of bureaucracy in Wikipedia~\cite{shaw2014laboratories}. Lastly, the FAQ length, reflecting established norms from collective discourse, also shows sublinear scaling. This highlights how coordination shapes norms, culture, and rules, and reveals that FAQ length grows even slower than the number of administrators.

Revisiting Fig.~\ref{fig:costs_vital}a, we have identified a plausible explanation for how complexity and authority influence the scaling factor of coordination costs. Let us revisit the phase diagram. According to the results, when the coordination mechanism deals with complexity but lacks authority (bottom left), $\beta$ has a high value ($\beta \simeq 1.3$, two-way coordination). Conversely, when the coordination mechanism exhibits similar complexity (top left) but relies on authority, $\beta$ can decrease ($\beta \simeq 0.9$, one-way coordination). Furthermore, when the coordination mechanism handles tasks of low complexity (top left), $\beta$ may even decrease further ($\beta \simeq 0.7$). Individuals can coordinate personally through communication and discussion~\cite{mintzberg1980structure, march1991simon}. Alternatively, rules and protocols can manage individuals impersonally~\cite{weber2023bureaucracy}. These coordination methods can complement and substitute each other to meet group coordination demands. Our findings reveal a shift in scale economy from superlinear to sublinear scaling, transitioning from personalized two-way coordination (bottom right) to personalized one-way, and finally to impersonal one-way coordination (top left) in Fig.~\ref{fig:costs_vital}a.

\section*{From Modular Interactions to Hierarchical Structure}

The superlinear relationship of two-way coordination suggests that as the number of contributors increases, the more interactions per person. Then, if every interaction requires administration, the number of administrators might also increase at the same rate $\approx N^{1.2}$. However, contrary to this expectation, the number of administrators increases sublinearly ($\approx N^{0.7}$) indicating economy of scale. This discrepancy raise an intriguing question: Where does this gap originate? Does the structure of two-way coordination reveal an economy of scale in one-way coordination, despite the lack of such economies in two-way interactions~\cite{clement2018searching}?

\begin{figure*}[!t]
    \centering
    \includegraphics[width=\textwidth]{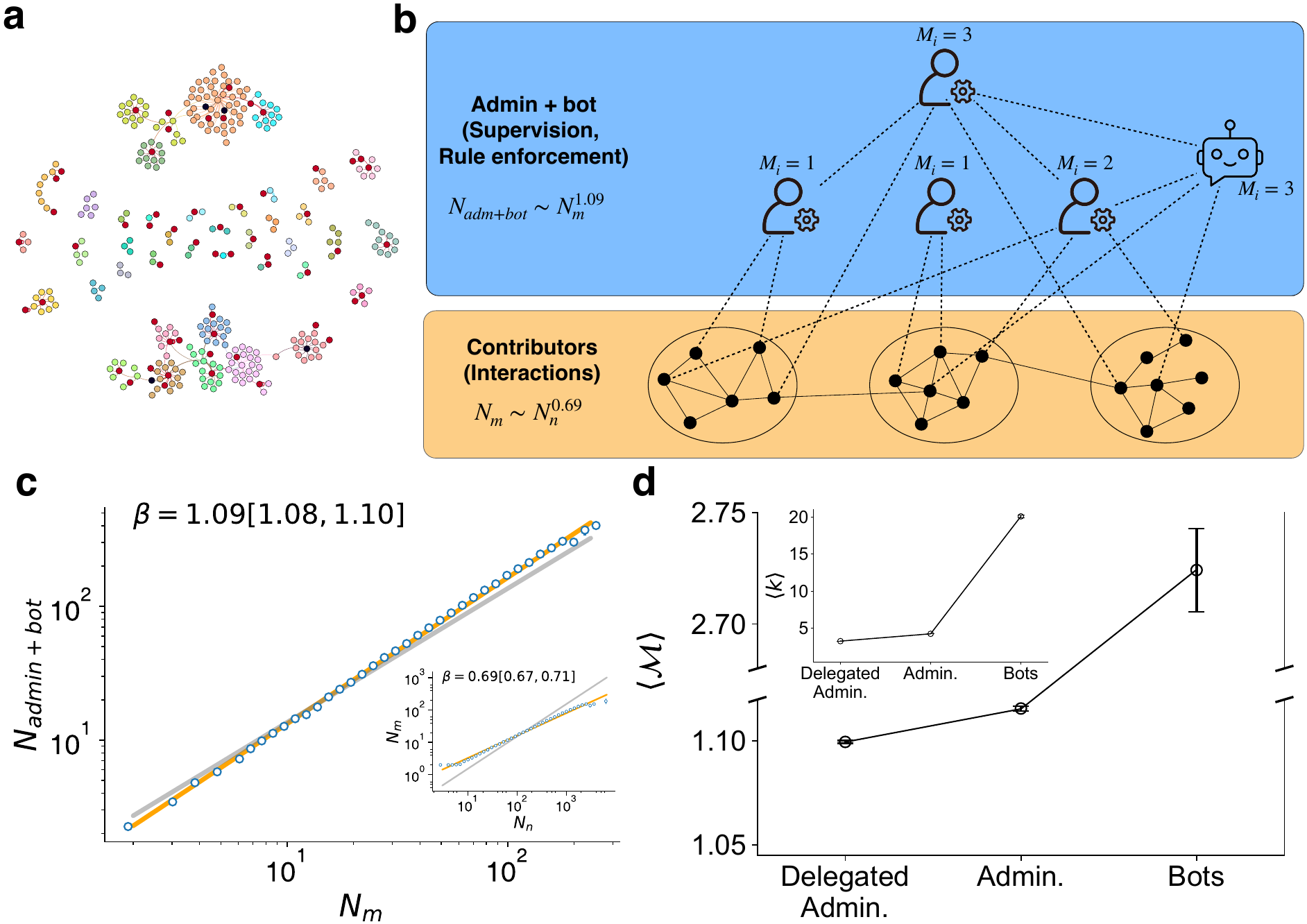}
    \caption{\textbf{Emergence of Formal Hierarchical Structures from Modular Interactions.} \textbf{(a)} Interaction network of the Phil Spector project. Node color indicates the extracted module. Red nodes represent administrators, and black nodes represent bots. \textbf{(b)} Schematic diagram of the interaction network. The modular structure naturally emerges from interactions, with administrators and bots managing these modules. The number of managed modules varies depending on the type of administrators and bots involved characterized by $\mathpzc{M}_i$ . \textbf{(c)} Scaling behavior of the number of administrators and bots relative to the number of modules. Each project is log-binned (blue) with error bars, and the estimated $\beta$ in $N^\beta$ (orange line) with a linear trend (gray line) for reference. The number of administrators and bots exhibits nearly linear scaling ($\beta$=1.09). The inset shows the scaling behavior of the number of modules relative to the number of nodes, demonstrating a sublinear scaling relationship. \textbf{(d)} The extent of module coverage for each role increases from delegated administrator, administrator (sysop) to bots (see Fig.~\ref{fig:adminship} in SI). The inset shows the average degree for each role.}
    \label{fig:module}
\end{figure*}

To understand this (dis-)economy of scales, we examine the interaction network (See Methods) from the mention relationship in the edit history. As an example, the interaction network of the Phil Spector project is shown in Fig.~\ref{fig:module}a. Overall, a modular structure naturally emerge from the interactions among contributors~\cite{brunswicker2021evolution}. Administrators (red nodes) and bots (black nodes) are broadly interacting across modules, holding the responsibility of managing these emergent modules. Building on this idea, we construct a schematic diagram of interaction networks in Fig.~\ref{fig:module}b. First, the modular structure naturally emerges from the mutual interaction among contributors without hierarchy. Second, administrators and bots try to regulate contributors utilizing emergent modules.

To test idea in Fig.~\ref{fig:module}b, we expand our analysis of interaction networks to encompass all vital articles. We define modules as those extracted from community detection (See Methods). Consistent with our observations in example networks, we find robust modular structures across all vital articles, with an average modularity score of $0.79$. As illustrated in Fig.~\ref{fig:module}c (inset), the number of modules $N_m$ increases sublinearly ($\beta=0.69$) as the number of nodes $N_n$ grows, originating from heterogeneous module sizes (Fig.~\ref{fig:mention}). Interestingly, Fig.~\ref{fig:module}c shows that the number of unique administrators and bots increases almost linearly ($\beta = 1.09$) with the number of modules, indicating $N_{\text{admin + bot}}\sim N_m$. This suggests that administrators and bots manage coarse-grained modules, dense groups of interactions, rather than individual interactions. 

Then, how do administrators and bots manage the modules? While one might initially assume a straightforward scenario where each admin manages approximately one module, we find a more nuanced structure emerges upon closer examination of the network~\cite{arazy2015functional}. In Fig.~\ref{fig:module}d, we address these questions with the average module coverage $\langle \mathpzc{M} \rangle$ (see Methods) for each administrative role. The $\langle \mathpzc{M} \rangle$ of delegated administrators is 1.10, whereas the  $\langle \mathpzc{M} \rangle$ of administrators, who hold higher authority than delegated administrators, is 1.12. This suggests that administrators with greater authority span a larger number of modules $\langle \mathpzc{M} \rangle$. On the other hand, Bots characterized by their high scalability and impersonal nature, exhibit the highest $\langle \mathpzc{M} \rangle=2.72$, engaging in broader modules rather than being central~\cite{gonzalez2021bots}. 

However, when we divide $\langle \mathpzc{M} \rangle$ by its degree $\langle k \rangle$ (which we define as the module awareness score $r$, see methods), we uncover different strategies for managing modules between human administrators and bots. $r$ ranges from zero to one. If 
$r$ is close to 1, connections are distributed across modules, indicating a module-aware strategy. If $r$ is close to 0, connections are randomly distributed neglecting modular structure, suggestion a individual-aware strategy. Our analysis shows that human administrators are more inclined towards a module-aware strategy ($r_{\text{del. adm.}}=0.34$, $r_{\text{adm.}}=0.26$), while bots tend to follow an individual-aware strategy ($r_{\text{bot}}=0.14$). In simper terms, the economy of scale in supervision of human administrators stems from the reduction processes originated from nascent modular structure, whereas the economy of scale in rule enforcement of bots comes from the extensive scalability. Our overarching findings are consistent even in the revert network: another types of interaction network (Fig.\ref{fig:revert_network}).


\section*{Emergence of Impersonal Coordination}
The scaling relationship generally describes behavior of coordination cost  with organization size $N$. However, some projects require unconventional coordination costs that deviate from the expected scaling law. As we discussed above, coordination costs are influenced not only by size $N$ but also by the complexity of functions $C$, which is often interrelated. In Wikipedia, projects with complex topics often require larger groups for a comprehensive perspective, leading to more disagreements, disputes, and extensive discourse. For instance, in categories like \textit{Everyday Life} and \textit{Mathematics}, the talk page size increases mildly superlinearly with contributors ($\beta \approx 1.1$, Table~\ref{table:costs}). In contrast, contentious topics like \textit{History} or \textit{Philosophy and Religion} exhibit a much steeper increase ($\beta \approx 1.5$, Table~\ref{table:costs}). Additionally, the relationship among diverse coordination mechanisms adds another layer of difficulty to fully understanding the coordination mechanisms at play. For instance, given organization size $N$, one might expect a trade-off between one-way coordination and two-way coordination: with more regulations, there might be fewer disputes or interactions. However, these questions has not been systematically tested yet.

 \begin{figure}[!t]
    \centering
    \includegraphics[width=1.0\textwidth]{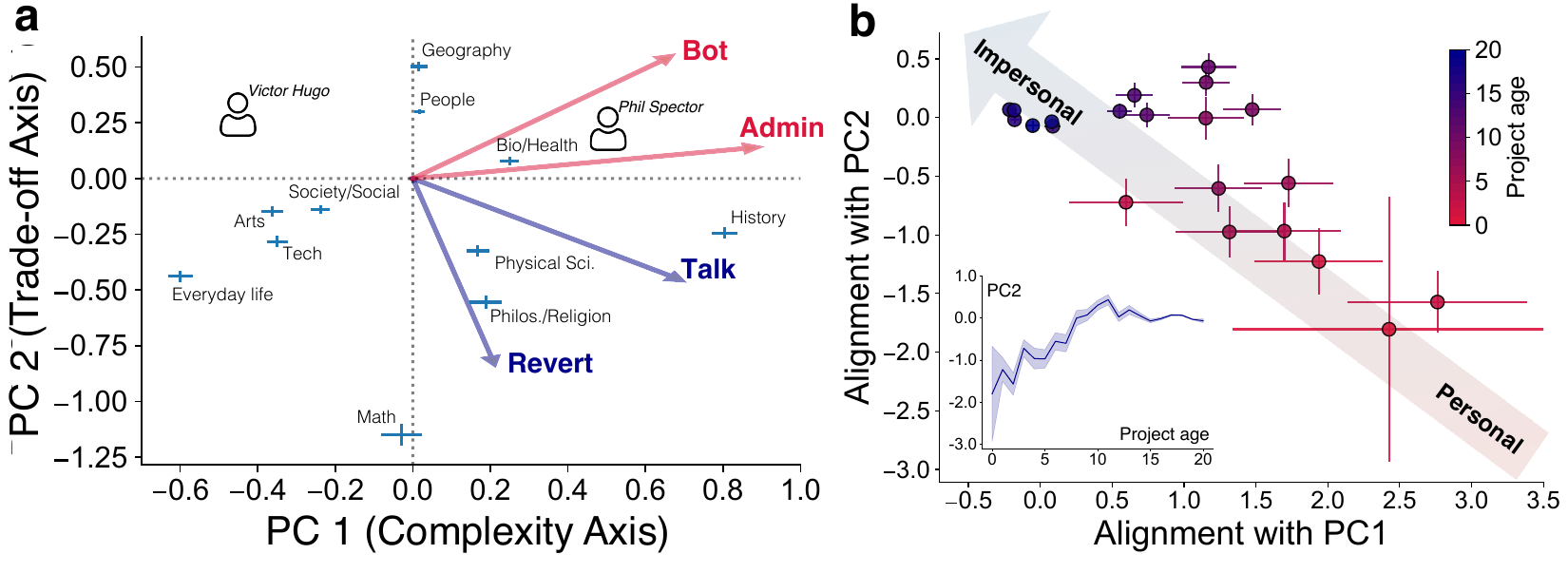}
    \caption{\textbf{Transition from Personal Coordination to Impersonal Coordination}
    \textbf{(a)} Compensation among coordination costs. Blue dots and whiskers indicate the average value of each principal axis categorized by a specific criterion with standard errors. The arrows in the plot indicate the loadings of each variable, showing how each original variable contributes to each principal component. Notably, all variables demonstrate positive loading for principal axis 1. However, for principal axis 2, we observe positive loading (indicated by a red arrow) and negative loading (represented by a blue arrow) for different variables. \textbf{(b)} Transition from personal coordination to impersonal coordination. Dots and whiskers indicate the average value of each principal axis categorized by a project year with standard errors. The time series of PC2 scores is depicted in the inset.
    }
    \label{fig:compensation}
\end{figure}

To understand factors influencing coordination costs beyond the average behavior of $N$, we quantify the residual of each coordination cost, which is difference between the observed and predicted values from scaling law, denoted as $\xi_i \equiv \log \frac{Y_i}{Y(N_i)} = \log \frac{Y_i}{Y_0N_i^{\beta}}$ ~\cite{bettencourt2010urban}. We find that administrative activities consistently correlate positively with other coordination activities, while other correlations are negligible ($<0.1$). This might suggests that governance, represented by administrative activities, plays a central role in the coordination process\cite{o2007emergence,demil2006neither}. Additionally, there is no trade-off mechanism, contradicting the common assumption that more regulations, fewer disputes. This lack of trade-off raises questions: Why is there no correlation? Could the strong correlation with administrative activities be confounding the dynamics within these residuals? For instance, controversial topics may drive coordination costs and administrative activities, resulting in a positive correlation with administrative activities but not among themselves.

To find hidden trade-off mechanism, we employ principal component analysis on the residuals (See Methods). Fig.~\ref{fig:compensation}a shows the two dominant patterns --- principal components. In PCA, \textit{loadings} indicate the correlation between original variables and principal components. High loadings, either positive or negative, suggest that specific variables strongly influence the respective principal component. First, We identify a ``complexity axis" (Principal Component 1, PC1), which shows an escalation of all coordination costs aligned with administrator activity, with a loading factor of 0.8, illustrated by the length of the arrows in Fig.~\ref{fig:compensation}a. Administrators typically focus on controversial projects, rarely facilitating transitions from high-conflict to low-conflict states~\cite{dedeo2016conflict}, leading to concurrent increases in all coordination costs. Consequently, contentious topics like \textit{History} and \textit{Philosophy/Religion} are positioned on the right (PC1 $>$ 0), while less contentious topics like \textit{Everyday Life} and \textit{Mathematics} are on the left (PC1 $<$ 0). For illustrative example, in the \textit{people} category, Phil Spector has a PC1 value of 0.53, indicating high contention, whereas Victor Hugo, with a similar number of contributors, has a PC1 value of -0.43.

In contrast to PC1, Principal Component 2 (PC2) reveals a hidden trade-off mechanism among coordination costs: one-way coordination (administrator and bot activity) with positive loadings suppresses two-way coordination (talk page and reverts) with negative loadings. Consequently, we refer to PC2 as the ``trade-off axis". Specifically, when focusing on coordination costs with high loadings, bot activity efficiently suppress conflicts through \emph{impersonal coordination} by rule, inherently mitigating conflicts and promoting smoother coordination~\cite{geiger2013levee, Brinkmann2023}. Aligned with the idea that PC2 is linked to impersonal coordination by bot activity, the PC2 scores indicate the effectiveness of rule-based coordination. Higher PC2 scores signify more effective coordination by impersonal coordination. In the inset of Fig. \ref{fig:compensation}b, we show PC2 scores grouped by project age, a proxy for article maturity. Early-stage projects (left) have negative PC2 scores, indicating weak impersonal coordination. As articles mature (right), PC2 scores increase, showing a transition from personal to impersonal coordination as Weber's theory of organizational evolution suggested \cite{weinberg1994}.

Taking into account the scores for PC1 and PC2 together, Fig.~\ref{fig:compensation}b illustrates the transition from personal coordination to impersonal coordination. In the early stages of a project, high PC1 and low PC2 scores indicate a reliance on personal coordination, with significant variance compared to older projects. As projects mature, average PC1 values decrease or stabilize, while PC2 values steadily increase, signaling the emergence of impersonal coordination. This shift aligns with organizational design theory: early-stage organizations use informal coordination, like mutual interactions, to explore and build new models under high uncertainty~\cite{burns1961management}. As they mature, they transition toward bureaucratization, adopting formal structures for stability and efficiency~\cite{desantola2017scaling, marchetti2022organizational}. Early contributors to nascent Wikipedia projects collaborate like startup founders, building initial outlines and fleshing out content. Facing disagreements, they engage in unstructured discussions and manually revert edits due to a lack of established rules. These mutual interactions help set the basic framework. As articles mature, coordination shifts to formal methods like bots and established norms or rules, since there are now enough prior cases for reference and most conflicts are minor. This transition to impersonal coordination is evident in mature articles but not in those with just a large number of contributors, as shown in Fig.~\ref{fig:real_size}.


To understand the trade-off between coordination mechanisms, we develop a minimal model (see Supporting Information). The parameter $\sigma$ governs the necessary regulation level based on project size; higher $\sigma$ values indicate greater deviations from expected scaling behavior. Each coordination mechanism then allocates a proportion of the required regulation following a predetermined compensation order. Using this model, we reproduced our empirical compensation results (Fig.\ref{fig:synthetic_combination}). Aligned with our explanation of PC1, as $\sigma$ rises, all coordination costs increase simultaneously. For PC2, using the compensation order ``bot-admin-talk-revert, we replicated results similar to empirical findings, suggesting this simple mechanism explains Wikipedia's coordination structure. Notably, this order matches the increasing pattern of scaling factors.


\section*{Mathematical Model of Coordination Mechanism  }

Motivated by the interaction structure, we propose a mathematical model for coordination mechanisms that lead to one-way coordination (administrator and bot), denoted as $C_1$, and two-way coordination (talk and revert), denoted as $C_2$, which explains our empirical findings. In formulating the model, we made two crucial assumptions. First, we assumed that contributors of the Wikipedia project only interact within in the naturally created nascent \textbf{module}, which is supported by our empirical observation. For instance, when contributors A and B are part of the same module, A's contribution to the project prompts a reaction from B if B disagrees with A's contribution. On the other hand, if A and B are in different modules, any action taken by A does not elicit a reaction from B. 

As a second assumption, contributors address conflicts through two-way coordination $C_2$ initially. With probability $p_e$, unresolved problems escalates to one-way coordination $C_1$, requiring attention from administrators or bots (Fig.~\ref{fig:modelSketch}). High $p_e$ necessitates significant amount of one-way coordination. Then, we assume that individuals learn from experience from past interaction within module, and this learning, denoted as $k$, accumulates within the organization. As accumulated learning $k$  increases, the probability of escalation decreases, following $p_e \propto 1/k$. In other words, organizational learning can reduce the need for one-way coordination.

  \begin{figure*}[t]
    \centering
    \includegraphics[width = 0.9\textwidth]{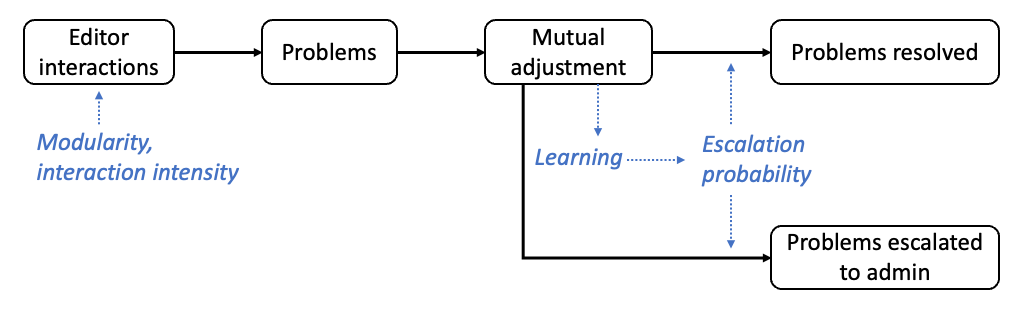}
    \caption{\textbf{Conceptual illustration of the model's variables and their relationships.} 
Problems arise from contributors' interactions and conflicts. Initially, contributors try to resolve these issues through two-way coordination. If unsuccessful, problems are escalated to admins. The escalation rate $p_e$ is determined by organizational learning, derived from two-way coordination.}
\label{fig:modelSketch}
\end{figure*}

According to our model, the scaling factor in the two-way coordination $\beta_{C_2}$ is  
 \begin{equation}
    \beta_{C_2} = \beta_{mod} (1- \alpha) + \alpha, 
    \label{eq:two_way}
\end{equation}
where $\beta_{\text{mod}}$ is the scaling factor of the number of modules, $N_{\text{m}} \sim N^{\beta_{\text{mod}}}$, and $\alpha$ is the interaction density within the module (See Supporting Information for the derivation). In our model, contributors interact only within their modules, so the total amount of two-way coordination $C_2 \sim N_m I_m$, where $N_m$ is the number of modules and $I_m$ is the number of interactions within a module. The number of modules $N_m$ increases with the number of contributors as $N_{\text{m}} \sim N^{\beta_{\text{mod}}}$. A $\beta_{\text{mod}} = 0$ suggests a constant number of modules regardless of project size, while a $\beta_{\text{mod}} = 1$ suggests that the number of modules increases linearly with the number of contributors, leading to a constant number of contributors per module. Empirically, $\beta_{\text{mod}}$ for the mention interaction network is 0.69, and this value is remarkably consistent across different types of interaction networks. Next, the number of interactions within a module $I_m$ is determined by its interaction density $\alpha$, within the module, $I_{\text{m}} \sim n^{\alpha}$, where $n$ is the number of contributors in the module. With \( 1 \leq \alpha \leq 2 \), this shows how the number of potential interactions scales with module size. An $\alpha$ value of 2 signifies a well-mixed interaction pattern within the module, while an $\alpha$ value of 1 suggests that each contributor interacts with a constant number of individuals, irrespective of module size. 

\begin{figure}[!t]
    \centering
    \includegraphics[width=1.0\textwidth]{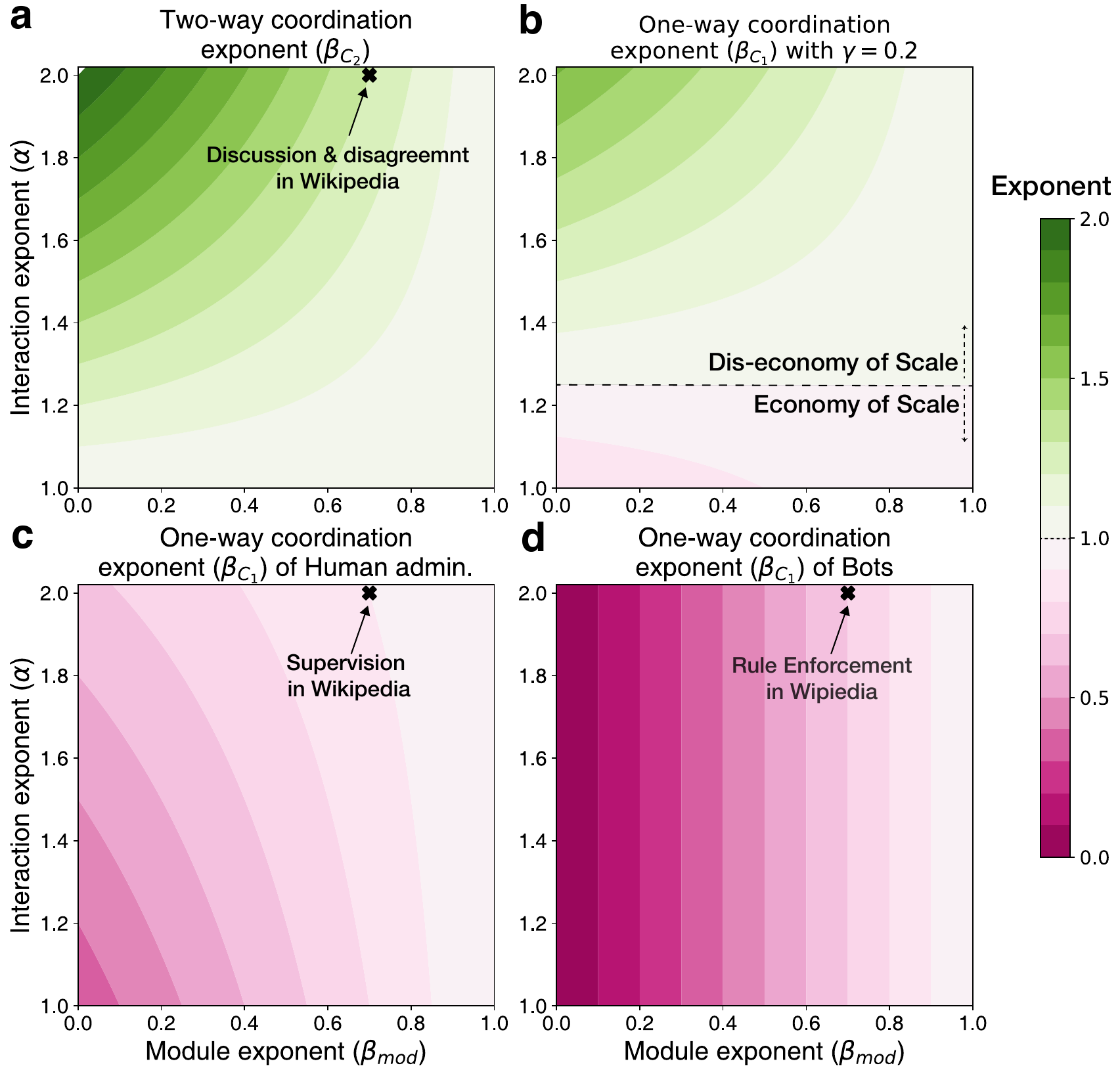}
    \caption{\textbf{Mathematical model of coordination mechanism.} Contour plots showing the model's predictions for scaling exponents as a function of the module scaling and interaction scaling exponents Color white denotes linear scaling. A deeper green color denotes more superlinear scaling, and a deeper magenta color denotes more sublinear scaling. \textbf{(a)} the two-way coordination exponent, $\beta_{C_2}$. 
    \textbf{(b)} the one-way coordination exponent ($\beta_{C_1}$) with $\gamma = 0.2$, \textbf{(c)} $\beta_{C_1}$ of human administrator ($\gamma = 2/3$), \textbf{(d)} $\beta_{C_1}$ of bot ($\gamma = 1$). Cross marker illustrates the status of discussion, disagreement, supervision, and rule enforcement on Wikipedia.}
    \label{fig:modelContour}
\end{figure}

 First, the model explains the inevitable nature of the superlinear scale of two-way coordination. Considering that $\beta_{\text{mod}}$ ranges from 0 to 1 and $\alpha$ ranges from 1 to 2, our model predicts $1 \leq \beta_{C_2} \leq 2$ (Fig.~\ref{fig:modelContour}a), where the minimum value of 1 with $\beta_{\text{mod}} = \alpha = 1$ and the maximum value of 2 with $\beta_{\text{mod}} = 0$ and $\alpha = 2$. This suggests that the scaling factor of two-way coordination is always greater than 1, indicating that a superlinear scale of two-way coordination is inevitable. Second, the model can inference the latent interaction density $\alpha$. Since the interaction network in Wikipedia includes various interactions, from strong forms like mentions and reverts to weak forms like co-editing, it is challenging to empirically examine the network comprehensively. Instead, we observe a plausible increasing pattern of $\alpha$ from strong to weak forms of interaction. Given the empirical scaling factor of two-way coordination in Wikipedia, $\beta_{C_2} = 1.3$, and $\beta_{\text{mod}} = 0.7$,  then we can infer $\alpha$ of underlying interaction network of Wikipedia is 2 (cross marker in Fig.~\ref{fig:modelContour}a).

Similarly, the scaling factor of one-way coordination $\beta_{C_1}$ is
 \begin{equation}
    \beta_{C_1}= \beta_{mod} + (1-\beta_{mod}) \alpha (1-\gamma),
\end{equation}
, where $\gamma$ is the learning rate (See Supporting Information for derivation). As mentioned above, an organization acquires knowledge $k$ from resolving past disputes, which is reflected in established norms, consensus, or guidelines. According to our first assumption, learning only happens within the module, resulting in $k \sim (I_{\text{m}})^{\gamma}$, where $0 \leq \gamma \leq 1$ represents the learning rate. A $\gamma$ value of 0 implies no learning, while a $\gamma$ value of 1 suggests perfect learning, meaning all experiences are learned. Higher $\gamma$ leads to a greater accumulation of experiences $k$, resulting in a lower probability of problem escalation and a reduced demand for one-way coordination.

 First, our model explains that economies of scale are not guaranteed and are only possible with high organizational learning. Fig.~\ref{fig:modelContour}b shows the scaling factor of one-way coordination, $\beta_{C_1}$, for $\gamma=0.2$, varying with $\beta_{mod}$ and $\alpha$. Notably, $\beta_{C_1}$ exhibits distinct behaviors based on the learning rate $\gamma$. With low organizational learning $k$ economies of scale in one-way coordination ($\beta_{C_1}<1$) are not guaranteed for all parameter combinations. In Fig.~\ref{fig:modelContour}b, regardless of $\beta_{mod}$, dis-economies of scale (green) and economies of scale (red) coexist, depending on the interaction exponent. Higher interaction density leads to dis-economies of scale in one-way coordination. Second, since we already know the empirical values of $\beta_{C_1}$, $\beta_{mod}$, and $\alpha$, the model can infer the latent variable $\gamma$. Fig.~\ref{fig:modelContour}c shows the scaling factor of human administrators with $\gamma=2/3$, and Fig.\ref{fig
}d shows the scaling factor of bots $\beta_{C_1}$ with $\gamma=1$. In both cases, a high learning rate 
$k$ ensures economies of scale, suggesting that the Wikipedia system could be sustainable and highlighting the importance of organizational learning in coordination processes.

The model also predicts the critical value $\gamma_c$ when economies of scale in one-way coordination happen. When $\gamma > \gamma_c = (\alpha - 1)/\alpha$, the scaling factor $\beta_{C_1} < 1$, indicating an economy of scale. Notably, the critical learning rate $\gamma_c$ is an increasing function of $\alpha$, suggesting more intensive interactions within a module (high $\alpha$), a higher learning rate $\gamma$ is required to achieve an economy of scale in one-way coordination. The model predicts the lower limit of $\beta_{C_1}$ when $\gamma = 1$, as in the case of bots (Fig.Fig.\ref{fig:modelContour}d), where $\beta_{C_1}$ equals $\beta_{\text{mod}}$, suggesting that rule enforcement in Wikipedia is near optimal. Lastly, the model explains the empirical findings on residuals (Tab.~\ref{table:resi}, See Supporting Information for detail)

\section*{Discussion}

Our study contributes to understanding how collective actions transcend the sum of individual efforts, especially in self-organizing systems characterized by open, voluntary participation. Specifically, first, we explore and measure the coordination mechanisms inherent in self-organized systems, highlighting the concept of a `minimum bound' of coordination cost emerging from self-organizing nature. Second, our paper reveals two distinct types of one-way coordination and two-way coordination through the lens of scaling factors, integrating them into a unified framework. Third, we emphasize the significance of transitioning from personal coordination to impersonal coordination, which has profound implications for the evolution of organizational structures. Fourth, we propose a comprehensive model that synthesizes these observations by leveraging learning mechanisms through interactions.

If organizations exist due to coordination action~\cite{chandler1978visible, galbraith1974organization}, transaction costs~\cite{coase1995nature, williamson1981economics}, and resource ownership~\cite{barney1986strategic, penrose1999growth}.  As technological advancements reduce coordination and transaction costs, the structure and dynamics of organizations may evolve, becoming less rigid and static. Our research aims to explore this phenomenon by expanding and redefining the concept of organizations. We break down actions into distinct components and assemble them into a typology of coordination mechanisms, which challenges our thoughts on the limits of organization.

This study also have several limitations. First, we employee the integral approach of measuring organization size and associated coordination costs. While our integral approach provides a comprehensive overview, it may overlook short-term fluctuations and dynamic changes within the organization. Future research could benefit from incorporating temporal data to analyze how coordination costs evolve over time. Second, we focused exclusively on self-organized systems, Wikipedia. It would be interesting to investigate top-down systems, such as companies or governments, and compare the results across diverse organizations. Examining these different organizational structures (e.g. top-down vs bottom-up) could enhance our understanding on coordination mechanism and help validate the model's applicability in various contexts. By understanding how coordination costs and mechanisms differ between self-organized and top-down systems, we can gain deeper insights into optimizing coordination in a wide range of settings. Lastly, our model assumes homogeneity in learning rates and interaction densities, which may not accurately reflect the diversity within different projects. Exploring heterogeneity in these parameters could provide a more nuanced understanding of coordination dynamics.

Our findings contribute to the literature on coordination costs and scaling in complex systems by shedding light on the challenges faced by online communities. People often overlook the costs of collective intelligence, assuming the absence of a formal organizational structure. However, in line with the natural law of hierarchy~\cite{hyland1995democratic, shaw2014laboratories}, we reveal the emergence of hierarchical structures and impersonal coordination in Wikipedia. These results have practical implications for online communities, highlighting the importance of understanding the coordination costs associated with different content categories and the need to develop appropriate coordination mechanisms as community size increases.

In our increasingly complex society, the efforts and expertise of highly specialized individuals are distributed across the globe. This necessitates their productive coordination across vast multi-level networks to create a diverse array of complex goods and services \cite{hosseinioun2023deconstructing}. The impact of successful collaboration is becoming increasingly important in knowledge production, highlighting the significance of collaborative efforts in producing successful outcomes and advancing scientific progress~\cite{Vanderwouden2023, wuchty2007increasing, fortunato2018science, wu2019large}. This reality underscores the urgent need for a deeper understanding of coordination mechanisms and their latent structures at various scales.

\section*{Materials and Methods}

\subsection*{Wikipedia Data}
We used the Wikipedia XML dump of English Wikipedia on January 9, 2022, comprising a history of edits of 6,412,821 different pages in Wikipedia. Each edit includes a unique identifier of the editor (username for registered contributors and IP address for unregistered contributors), timestamp, comments, and edit size. Additionally, we extracted contributor's roles from the user group table of the SQL dump. We focus our empirical analysis on Vital Articles, a subset of projects selected by Wikipedia that cover important topics, with the aim of higher quality.  
These articles have been carefully selected to cover a broad spectrum of subjects and are considered essential for obtaining a comprehensive understanding of human knowledge. Encompassing a wide range of disciplines, including history, science, arts, geography, and more, these articles represent significant topics, events, individuals, and concepts. They undergo regular evaluation and updating to ensure their accuracy, relevance, and comprehensiveness, thus serving as indispensable pillars of knowledge within Wikipedia's vast and extensive encyclopedia. Among them, we only consider projects with at least one edit in their talk pages, one revert in their content, and one action from an administrator. This resulting subset comprises 26,014 projects tuned to our questions.

\subsection*{Scaling Analysis}

To study how coordination costs increase with the number of contributors, we use the scaling framework~\cite{west2018scale}. Using the number of unique contributors on each page $N$, as the measure of organization size, we consider power-law scaling to take the form of 
\begin{equation}
    Y = Y_0N^{\beta}.
    \label{eq:scaling_relation}
\end{equation}
$Y$ denotes the coordination cost, while $Y_0$ serves as a normalization constant. The parameter $\beta$ characterizes the rate at which $Y$ increases in response to relative changes in $N$, with values of $\beta$ exceeding 1 indicating a faster increase and values below 1 indicating a slower increase compared to expected if it were linear. Here, we use a specific strategy to fit the data into the scaling law originated to focus on understanding growth behavior, which is the interest of this research. In our data set, the majority of projects have a relatively small number of contributors, typically around 1,000. However, there are a few exceptional projects with a significantly larger contributor base. If we were to fit a model using ordinary least squares (OLS) regression, the model would naturally prioritize explaining the majority of the data. However, in this paper, our focus is on understanding the growth patterns of these projects, rather than modeling the majority of the data. To address this, we begin by logarithmically binning the data and fitting the model using the binned data. This approach ensures that every scale, from small to large projects, is given equal importance in our analysis. 

The exponents are robust to entire English Wikipedia articles beyond vital articles with 
$\beta_{\text{talk}}=1.38$, $\beta_{\text{revert}}=1.24$, $\beta_{\text{adm. act.}}=0.91$, and $\beta_{\text{bot act.}}=0.68$.
Also, among the 48,864 vital pages on Wikipedia, 22,680 pages have not undergone any administrative actions. To delve deeper into this matter, we performed a robust check by re-conducting a scaling analysis on talk page size, number of reverts, and bots' activity, including pages that have not received any intervention from administrators.  We again confirm that our results are consistent (Fig.~\ref{fig:costs_robust}), affirming our earlier observations.

\subsection*{Interaction Network}
To understand this (dis-)economy of scales , we construct the \emph{interaction network} from the edit history when contributors mention other contributors explicitly (e.g., \textit{Undid revision by @Jisung}). In this network, each node is the contributor who has mentioned another person or was mentioned by another person at least once. Nodes are linked when there exist mentions. For simplicity, we ignore directionality in this network. As almost 92\% of mentions are revert actions, the constructed mention network would not be too different from the revert network. For the analysis, 
we remove components (subgraph) with two nodes, which is a dyadic relationship. We extracted the modular structure with the community detection ~\cite{blondel2008fast}. In general, the constructed network is modular (the average modularity is 0.79).

\subsection*{Module coverage and module-awareness}
To quantify the extent of module coverage of each node, we use module coverage indicated as $\mathpzc{M}_i$, which is quantified by the unique number of modules managed by a given node $i$, which serves as a proxy for its role as a broker or structural hole~\cite{lazega1995structural}. As an illustration, in Fig.~\ref{fig:module}b, the leftmost administrator has two connections, but both connections are within a single module, resulting in $\mathpzc{M}_i = 1$. On the other hand, the top administrator has four connections spanning three unique modules, yielding $\mathpzc{M}_i = 3$. 
We calculate the average module coverage capacity for delegated administrators, administrators, and bots,  yielding $\langle \mathpzc{M} \rangle$ for each type. When we consider the concept of module coverage, the value $\langle \mathpzc{M} \rangle$ tends to increase as the average degree is high. For example, bots often exhibit a high $\langle \mathpzc{M} \rangle$ due to their elevated average degree. To measure module coverage behavior while normalizing for degree size, we utilize the module awareness ratio $r = \frac{\langle \mathpzc{M} \rangle}{\langle k \rangle}$ for delegated administrators, administrators, and bots. The value of $r$ can vary depending on the node's strategy for managing connections within the network. 

For better understanding, we propose two extreme cases. The first extreme scenario is the perfect module-aware strategy, wherein nodes establish connections solely with the module and not with individual nodes. In this case, multiple connections within the same module are not allowed, resulting in $\mathpzc{M}_{i} \simeq k_i$ and a value of $r \simeq 1$. Conversely, the other extreme scenario is the perfect individual-aware strategy, where nodes establish connections with individual nodes without considering the modular structure. In many large networks, it is theoretically possible for the size of the largest module to increase proportionally to the size of the entire network~\cite{newman2001random}, while the sizes of smaller modules remain constant. If nodes adopt the individual-aware strategy, connections are more likely to be distributed in the larger module. Consequently, $\mathpzc{M}_{i}$ will remain constant as $k_i$ increases, resulting in a value of $r \simeq 0$. Likewise, we can map the node's strategy in the network along a continuous spectrum ranging from the individual-aware strategy ($r=0$) to the module-aware strategy ($r=1$)

\subsection*{Residual}

As the scaling relationship (Eq.~\ref{eq:scaling_relation}) only provides an average description of the behavior of coordination costs, differences between observed and predicted values provide insight into the factors that affect coordination costs beyond the mean behavior predicted by the scaling law. To investigate this, we calculate the residual, $\xi_i$~\cite{bettencourt2010urban}, for each coordination cost as follows:

\begin{equation}\label{eq:resi}
    \xi_i\equiv\log \frac{Y_i}{Y(N_i)} = \log \frac{Y_i}{Y_0N_i^{\beta}},
\end{equation}
where $Y_i$ is the observed value of coordination costs and $Y(N_i)$ is the expected value of coordination costs given the number of contributors $N_i$. For the comparison, Table~\ref{table:raw}, presents the Pearson correlation between coordination costs, which exhibits strong correlations primarily influenced by the number of contributors. On the other hand, 
Table~\ref{table:resi} shows the correlation among residuals, highlighting their relationships after removing the average description of the behavior of coordination costs.

\subsection*{Principal component analysis}

Principal Component Analysis~\cite{pearson1901liii}, commonly known as PCA, is a powerful statistical technique used for dimensionality reduction and data compression. PCA allows us to extract the most significant information from a dataset by capturing the underlying structure and reducing redundancy. PCA provides a streamlined and more manageable representation of data, facilitating better insights and analysis. In our analysis, we apply PCA to the residuals of each project as a four-dimensional vector, [$\xi^{\text{talk}}$, $\xi^{\text{revert}}$, $\xi^{\text{adm. act.}}$, $\xi^{\text{bot act.}}$], and extracted two principal components (PC1 and PC2). Explained variance ratio for Principal Component 1 (PC1) is 0.40, and for Principal Component 2 (PC2) is 0.23.

\section*{Acknowledgements}
The authors would like to acknowledge the support of the National Science Foundation Grant Award Number 2133863. J.Y and H.Y thank Eric Rupley for the generous support and interesting discussions.

\section*{Author Contributions}

J.Y. collected and performed analysis. J.Y. and H.Y. conceptualized, designed, and drafted the study. All authors interpreted the analyses, wrote the manuscript, and approved the final version of the manuscript. 
\section*{Additional Information}
Supporting Information is available for this paper. Correspondence and requests for materials should be addressed to Dr. Youn.

\section*{Data availability}
Wikipedia edit history data are available at wiki-dumps, \url{https://dumps.wikimedia.org/}. We collect the list of vital articles from \url{https://en.wikipedia.org/wiki/Wikipedia\%3AVital_articles}.
 
 \section*{Code Availability}
The code for this analysis is available upon request.

\bibliographystyle{naturemag}
\bibliography{main}

\pagebreak
\begin{center}
\textbf{\Huge Supplementary Information: What makes Individual \textit{I}'s a Collective \textit{We}; \\ Coordination mechanisms \& costs}
\end{center}
\setcounter{equation}{0}
\setcounter{figure}{0}
\setcounter{table}{0}
\makeatletter
\renewcommand{\theequation}{S\arabic{equation}}
\renewcommand{\thefigure}{S\arabic{figure}}

\section*{S1 Text. Who is the administrator in Wikipedia?: Bureaucracy in Wikipedia}
\label{si:text:whoisadmin}

\begin{figure*}[t]
    \centering
    \includegraphics[width=1.0\textwidth]{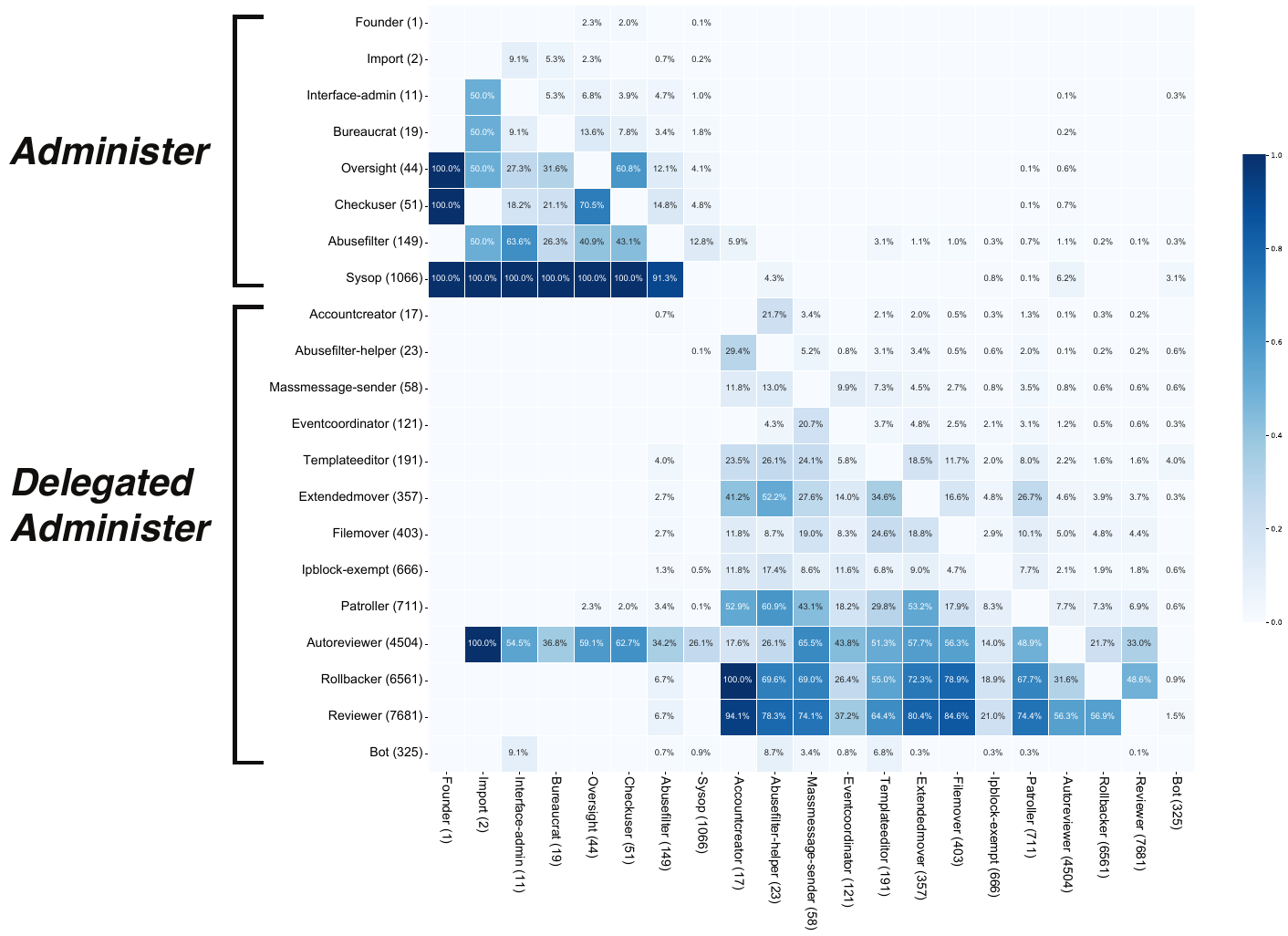}
    \caption{\textbf{A hierarchical structure of adminship flags in Wikipedia.} The axes of the figure represent lists of adminship flags. Since a user can possess multiple adminship flags, the colored squares with annotated numbers indicate the conditional probability that a user holding an adminship flag on the y-axis also holds an adminship flag on the x-axis. This representation reveals two distinct community structures within the adminship system: requested administrators and delegated administrators.} 
    \label{fig:adminship}
\end{figure*}

The level of authority granted to users on Wikipedia is determined by their user access level, which defines the actions they are allowed to perform. The hierarchical structure of adminship flags is depicted in Fig.~\ref{fig:adminship}. The most prominent type of administrator is known as a ``Sysop" (system operator), and they make up only 0.001\% of Wikipedia users. To become a sysop, individuals must undergo the adminship process and receive official approval, which grants them the authority to block and unblock users, protect and delete pages, and rename pages. Sysops also have the potential to acquire higher or more specialized access levels (Fig.\ref{fig:adminship} top). Due to the limited number of administrators, they often delegate specific responsibilities to trustworthy Wikipedia contributors (Fig.\ref{fig:adminship} bottom), referred delegated administrators). This category also includes bots, which are approved accounts used to assist human contributors in automating repetitive tasks. For analytical purposes, both requested, and delegated administrators are considered as ``administrators'' making up 0.003\% of Wikipedia users. For simplicity, we define an administrator as someone who has held the adminship flags at least once in Wikipedia's history.

\pagebreak

\section*{S2 Text. Mutual Interaction Network - Revert Network}
\label{si:text:revert_network}

\begin{figure*}[!t]
    \centering
    \includegraphics[width=\textwidth]{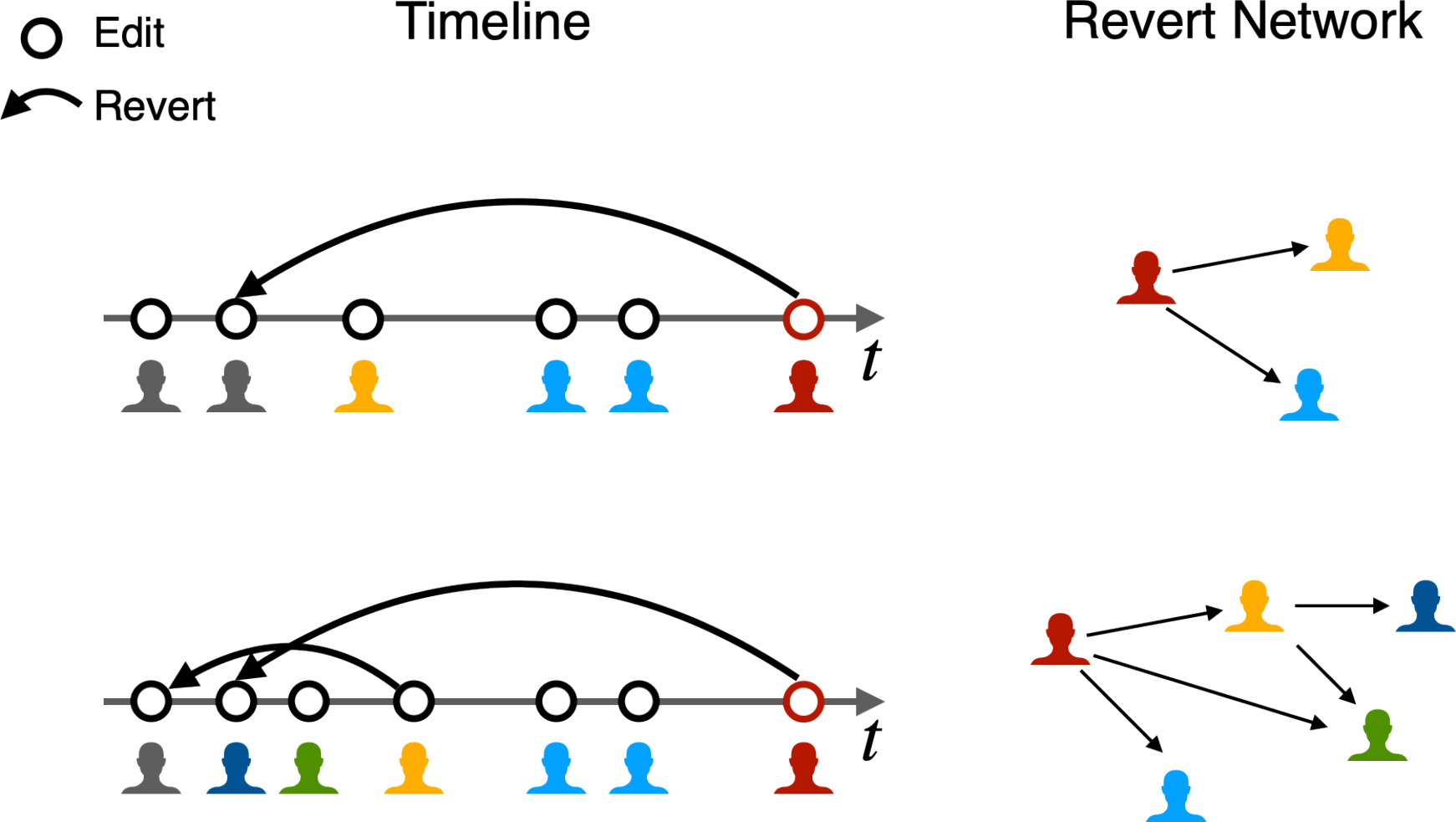}
    \caption{\textbf{Illustrative example of the revert network} (Top) In the scenario, the red contributor reverts the page back to the version contributed by the gray contributor, disregarding the edits made by the blue and yellow contributors. Consequently, in the reversion network,  red-yellow and red-blue links are established, each assigned a weight of $1/2$, calculated as one divided by the number of contributors reverted in each revert. (Bottom) Similarly, in a more intricate scenario, yellow contributor ignore the green and blue contributor and red contributor ignore blue, yellow, and green contributor. Consequently, yellow-green, yellow-blue links with weight $1/2$ and red-blue, red-yellow, and red-green links with weight $1/3$ are constructed.}
    \label{fig:revert_network_schem}
\end{figure*}

For robustness check, we construct an alternative mutual interaction network based on revert. In this network, each node represents a contributor who has either reverted another contributor's edit or been reverted by others at least once. Nodes are connected if one contributor reverts another's edit.  For instance, in Fig.~\ref{fig:revert_network_schem} (top), the red contributor reverts the project to a version four iterations behind, disregarding the contributions from the blue and yellow contributors. Consequently,  red-yellow and red-blue links are formed with a weight of $1/2$, calculated as one divided by the number of contributors reverted in each revert. For simplicity, we do not consider the directionality of these connections. Similar to mention network in the main text, we filter out components (subgraphs) comprising only two nodes, indicative of a dyadic relationship and extract modular structures using community detection techniques~\cite{blondel2008fast}.

\begin{figure*}[!t]
    \centering
    \includegraphics[width=\textwidth]{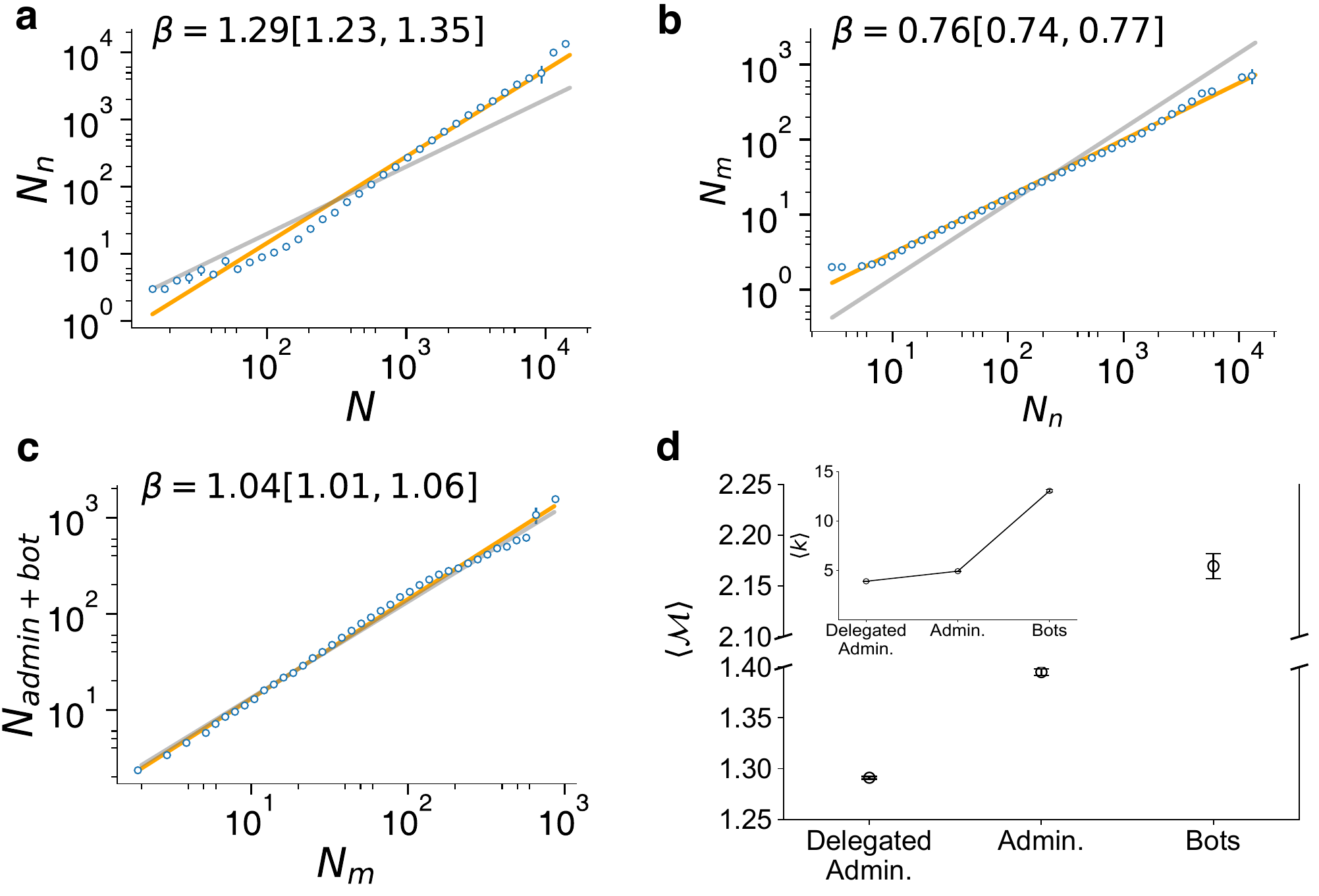}
    \caption{\textbf{Emergence of Formal Hierarchical Structures from Modular Interactions. --- revert network} \textbf{(a)}  Scaling behavior of the number of contributors and contributors in revert network. Each project is log-binned (blue) with error bars, and estimated $\beta$ in $N^\beta$ (the orange line), accompanied by a linear trend (gray line) for reference. \textbf{(b)} The number of modules shows a sublinear scaling relationship with the number of contributors in revert network.\textbf{(c)} Scaling behavior of the number of administrators and bots compared to the number of modules. The number of administrators and bots shows an almost linear scaling ($\beta$=1.04).  \textbf{(d)} The extent of module coverage for each role, increasing from delegated administrator to administrator (sysop) to Wikibots (see Fig.\ref{fig:adminship} in SI). The average degree of each role is shown in the inset. The overall pattern is remarkably consistent with the pattern in the mention network.
    }
    \label{fig:revert_network}
\end{figure*}

We discovered a parallel outcome to that shown in Figure \ref{fig:module} within the revert network (Fig. \ref{fig:revert_network}). First, we observed a robust modular structure similar to that of the mention network,with an average modularity score of 0.80. Second, As depicted in Fig. \ref{fig:revert_network}b, the number of modules $N_m$ exhibited sublinear growth ($\beta=0.76$) as the number of nodes $N_n$ increased. Third, Fig. \ref{fig:revert_network}c demonstrated that the unique count of administrators and bots increased almost linearly ($\beta = 1.04$) with the number of modules in the revert network, indicating $N_{\text{admin + bot}}\sim N_m$. Lastly, in Fig. \ref{fig:revert_network}d, we observed that the efficiency gains in supervision of human administrators stemmed from the module-aware capabilities, whereas the scalability in rule enforcement was facilitated by the extensive scalability of bots.

\pagebreak
\section*{S3 Text. Computational model explaining PCA analysis }

For a deeper understanding of the trade-off between the coordination mechanism and suggesting possible mechanisms behind it, we build the minimal model for generating synthetic data. In empiric, we have four coordination mechanisms, resulting in $4!=24$ possible compensation orders. For each compensation order (e.g. bot-admin-talk-revert), we generate the 10,000 synthetic data points (or page). For generating each data point, which is the synthesized project.

\begin{enumerate} 
    \item  Set the system size, denoted as $N_i$.
    \item Set the total amount of regulation required, denoted as $R_i$, using the formula $R_i := e^{x_i}$. The value of $x_i$ is generated from a Gaussian distribution $\mathcal{N}(0,\sigma^2)$, where we set $\sigma=1$. 
    \item Distribute the $R_i$ to the four dimensions of the compensation order for coordination mechanisms (e.g. bot-admin-talk-revert), resulting four dimensional vector $r^{\text{first}}$, $r^{\text{second}}$, $r^{\text{third}}$, and $r^{\text{fourth}}$ as follows
    \begin{itemize}
    \item Set value for $r^{\text{first}}_i$ from uniform distribution $\unif_{[0,R_i]}$
    \item Set value for $r^{\text{second}}_i$ from uniform distribution $\unif_{[0,R_i - r^{\text{first}}_i]}$
    \item Set value for $r^{\text{third}}_i$ from uniform distribution $\unif_{[0,R_i - r^{\text{first}}_i - r^{\text{second}}_i]}$
    \item Set value for $r^{\text{fourth}}_i$ from uniform distribution $\unif_{[0,R - r^{\text{first}}_i - r^{\text{second}}_i - r^{\text{third}}_i]}$
  \end{itemize}
  \item Finally, multiplying the scaling effect of contributor size to each dimension,\\ $Y_i:= [r^{\text{first}}_i N_i^{\beta_{\text{first}}}, r^{\text{second}}_i N_i^{\beta_{\text{second}}}, r^{\text{third}}_i N_i^{\beta_{\text{third}}}, r^{\text{fourt}}_i N_i^{\beta_{\text{fourth}}}$]. For scaling exponent beta, we uses the values from empiric.
\end{enumerate}

Here, we introduce two simple mechanisms to comprehend the empirical observations. First, let's consider $R_i= e^{x_i}$, where $x_i$ follows a Gaussian distribution with mean 0 and variance $\sigma^2$ which is the deviations from the average behavior. A high value of $R_i$ indicates that a page requires significant regulation relative to its size. In practical terms, this may be associated with the contentiousness of each page and the first principal component (PC1) in our PCA analysis. Second, sum of coordination mechanisms is approximately conserved, that is $\sum_i r_i = R_i$, and order of compensation (e.g. bot-admin-talk-revert) can bring insight of the hidden compensation patterns among compensation mechanism. When the initially chosen coordination mechanism consumes a significant portion of $R_i$, the remaining resources allocated to other coordination mechanisms become marginal. This phenomenon can elucidate the increasing order of loadings observed in the second principal component (PC2).

\begin{figure}[!t]
    \centering
    \includegraphics[width=\textwidth]{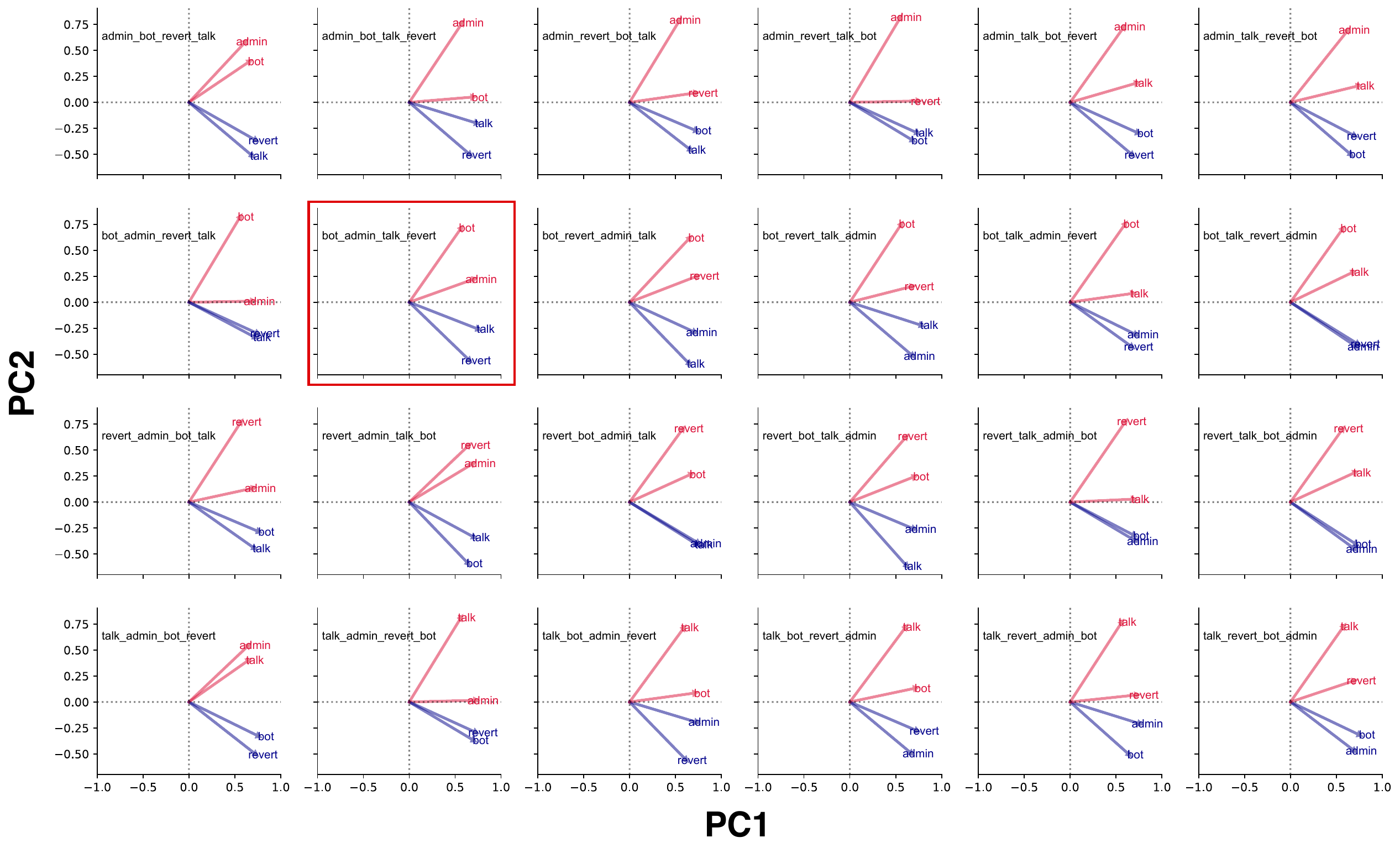}
    \caption{\textbf{PCA analysis on the synthetic data with compensation} We experiment with possible order of compensation for synthetic data. We found that the order of bot-admin-talk-revert (highlighted with a red square) is the most plausible compensation order that supports our empirical observation.
    }
    \label{fig:synthetic_combination}
\end{figure}

From our suggested model, we generate the synthetic data for possible compensation order, resulting in 24 distinct synthetic datasets. Similarly, we calculate the residual and decompose the residuals into two principal components. Results are shown in \ref{fig:synthetic_combination}. Surprisingly, when we set $\sigma=1$. the explained variance ratio for the Principal Component from synthetic data is very close to the explained variance ratio from empiric. The average explained variance ratio of synthetic data for Principal Component 1 (PC1) is 0.45, and for Principal Component 2 (PC2) is 0.21. We also found that the order of the ``bot-admin-talk-revert'' most explains the empiric, suggesting a hidden order of coordination in Wikipedia.

\section*{S4 Text: Mathematical model of coordination mechanism: Overview }

\begin{figure*}[t]
    \centering
    \includegraphics[width = 1\textwidth]{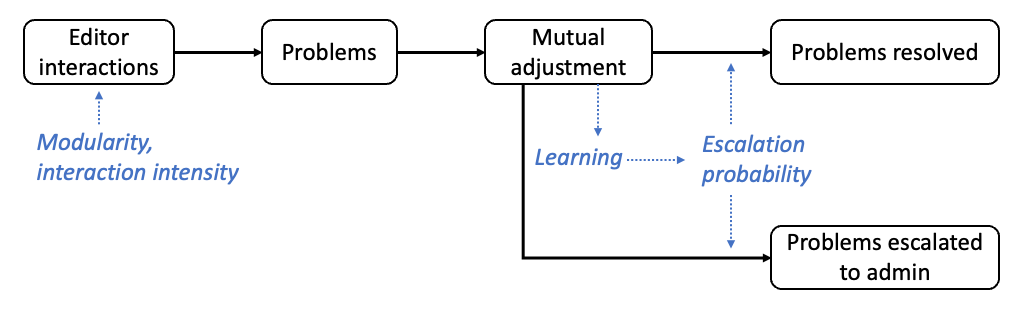}
    \caption{Conceptual illustration of the model's variables and their relationships. Arrows denote hypothesized causal relationship, the signs + and - denote whether the two variables are positively and negatively linked, respectively.  }
\label{fig:modelSketch}
\end{figure*}

 First, the model explains the inevitable nature of the superlinear scale of two-way coordination. (See Supporting Information for the derivation). As the first assumption, contributors only interact within the module. Then, the total interaction of given project $I \sim N_m I_m$ where $N_m$ is the number of modules and $I_m$ is the number of interactions within the module. Here, number of modules $N_m$ increases as number of contributor increases with $N_{\text{m}} \sim N^{\beta_{\text{mod}}}$. $\beta_{mod} = 0$ suggests the constant number of modules regardless of the project size, and $\beta_{mod} = 1$ suggests that the number of modules increases linearly with the module size, leading to a constant number of contributors in a module. The number of interactions within module $I_m$ is determined by its interaction density $\alpha$ within module $I_{\text{m}} \sim n^{\alpha}$ where $n$ is the number of contributors in the module. With \( 1 \leq \alpha \leq 2 \), gives how the number of potential interactions scale with module size. An $\alpha$ value of 2 signifies a well-mixed interaction pattern within the module, while an $\alpha$ value of 1 suggests that each contributor interacts with a constant number of individuals, irrespective of module size.  Subsequently, the overall interaction $I$, representing the necessary amount of two-way coordination, can be expressed as $I=N^{\beta_{\text{mod}} (1 - \alpha) + \alpha}$ (See Supporting Information). Building on this, the scaling factor of the two-way coordination is

 \begin{equation}
    \beta_{C_2} = \beta_{mod} (1- \alpha) + \alpha.
\end{equation}
Considering that $\beta_{\text{mod}}$ ranges from 0 to 1 and $\alpha$ ranges from 1 to 2, model suggest $1\leq \beta_2 \leq 2$ as depicted in Fig.\ref{fig:modelContour}. The minimum value of 1 is attained when $\beta_{\text{mod}} = \alpha = 1$, while the maximum value of 2 is reached when $\beta_{\text{mod}} = 0$ and $\alpha = 2$. In empirical findings related to the scaling factor of reverts and talk pages, $\beta_{C_2} = 1.3$, falling within this specified range. Our model posits that the superlinear scaling of bi-directional coordination should consistently be observed.

Next, our model elaborates on the economy of scale in one-way coordination. As previously mentioned, contributors acquire knowledge through experience, and this accumulated learning, denoted as $k$, is reflected in established norms, consensus, guidelines, or precedents for resolving past disputes. We assume that learning $k$ originates from interactions within a module, specifically $k \sim (I_{\text{m}})^{\gamma}$, where $0 \leq \gamma \leq 1$ represents the learning rate. A $\gamma$ value of 0 implies no learning, while a $\gamma$ value of 1 suggests perfect learning, meaning all experiences are learned. Then, the probability of escalating, $p_e\propto1/k$ is inversely proportional to the amount of learning. Then, total number of escalation is $N_{\text{e}} \sim N_{\text{m}} I_{\text{m}}  p_{\text{e}}$, which is the product of total interaction $I$ and probability of escalation $p_e$. Subsequently, the overall escalation $N_{\text{e}}$, representing the required amount of one-way coordination, can be expressed as $N_e \sim N^{\beta_{mod}+ (1-\beta_{mod})\alpha(1-\gamma)}.$. Then, the scaling factor of one-way coordination is

 \begin{equation}
    \beta_{C_1}= \beta_{mod} + (1-\beta_{mod}) \alpha (1-\gamma).
\end{equation}
Fig.~\ref{fig:modelContour} b,c shows the scaling factor of one-way coordination $\beta_{C_1}$ varied with $\beta_{mod}$ and $\alpha$. Notably, $\beta_{C_1}$ exhibits distinct behaviors based on the learning rate $\gamma$, and the economy of scale in one-way coordination ($\beta_{C_1}<1$) is not guaranteed for all parameter combinations. For instance, in the case of low $\gamma$ (Fig.\ref{fig:modelContour}b), there exist regions in the parameter space where the predicted exponent is super-linear. Conversely, when the $\gamma$ is high (Fig.\ref{fig:modelContour}b), an economy of scale in one-way coordination is consistently assured, underscoring the significance of organizational learning in coordination processes. 
The model also predicts the critical value $\gamma_c$ when the economy of scale in one-way coordination happen. When $\gamma > \gamma_c = (\alpha - 1)/\alpha$, the scaling factor $\beta_{C_1} < 1$, indicating an economy of scale. It is noteworthy that the critical learning rate, $\gamma_c$, is an increasing function of $\alpha$. This suggests that when there are more intensive interactions within a module (high $\alpha$), a higher learning rate $\gamma$ is required to achieve an economy of scale in one-way coordination. Lastly, the model predicts the lower limit of $\beta_{C_1}$ when $\gamma = 1$, signifying perfect learning, $\beta_{C_1}$ equals $\beta_{\text{mod}}$. Consequently, the maximum potential for economy of scale is determined by the modules within the interaction structure.

We have developed a mechanistic model for the two types of coordination mechanisms for decentralized projects like Wikipedia, depicted in \ref{fig:modelSketch}. The model is based on two key considerations.  (i) We consider the amount of two-way  coordination the result of the modularity and the intensity of interactions within modules. (ii) We consider one-way coordination in decentralized projects is to deal with escalated two-way interactions, and escalation rate decreases as editors learn from experience. Consideration (i) alone predicts the superlinear scaling in bi-directional coordination, such as talk page length and reverts. It also predicts that the scaling exponent is greatest for high interaction intensity and constant number of modules as the project grows. The scaling exponent is the smallest for low interaction intensity and number of modules linearly growing with project size. 

(i) and (ii) combined to predict that the scaling exponent for one-way coordination, such as administrator activity, can be superlinear or sublinear depending on the learning rate. We find a critical learning rate above which we predict sublinear scaling of admin activity. The critical learning rate is positively associated with interaction intensity. So the sublinear scaling of admin activity, although we observe in Wikipedia, is not universally guarenteed. The model suggests we observe this efficiency because Wikipedia editors learn from experience at a high enough rate, such as through establishing norms, past consensus, guidelines, etc.

Next, our model elaborates on the economy of scale in one-way coordination. As previously mentioned, contributors acquire knowledge through experience, and this accumulated learning, denoted as $k$, is reflected in established norms, consensus, guidelines, or precedents for resolving past disputes. We assume that learning $k$ originates from interactions within a module, specifically $k \sim (I_{\text{m}})^{\gamma}$, where $0 \leq \gamma \leq 1$ represents the learning rate. A $\gamma$ value of 0 implies no learning, while a $\gamma$ value of 1 suggests perfect learning, meaning all experiences are learned. Then, the probability of escalating, $p_e\propto1/k$ is inversely proportional to the amount of learning. Then, total number of escalation is $N_{\text{e}} \sim N_{\text{m}} I_{\text{m}}  p_{\text{e}}$, which is the product of total interaction $I$ and probability of escalation $p_e$. Subsequently, the overall escalation $N_{\text{e}}$, representing the required amount of one-way coordination, can be expressed as $N_e \sim N^{\beta_{mod}+ (1-\beta_{mod})\alpha(1-\gamma)}.$. Then, the scaling factor of one-way coordination is

 \begin{equation}
    \beta_{C_1}= \beta_{mod} + (1-\beta_{mod}) \alpha (1-\gamma).
\end{equation}
Fig.~\ref{fig:modelContour} b,c shows the scaling factor of one-way coordination $\beta_{C_1}$ varied with $\beta_{mod}$ and $\alpha$. Notably, $\beta_{C_1}$ exhibits distinct behaviors based on the learning rate $\gamma$, and the economy of scale in one-way coordination ($\beta_{C_1}<1$) is not guaranteed for all parameter combinations. For instance, in the case of low $\gamma$ (Fig.\ref{fig:modelContour}b), there exist regions in the parameter space where the predicted exponent is super-linear. Conversely, when the $\gamma$ is high (Fig.\ref{fig:modelContour}b), an economy of scale in one-way coordination is consistently assured, underscoring the significance of organizational learning in coordination processes. 
The model also predicts the critical value $\gamma_c$ when the economy of scale in one-way coordination happen. When $\gamma > \gamma_c = (\alpha - 1)/\alpha$, the scaling factor $\beta_{C_1} < 1$, indicating an economy of scale. It is noteworthy that the critical learning rate, $\gamma_c$, is an increasing function of $\alpha$. This suggests that when there are more intensive interactions within a module (high $\alpha$), a higher learning rate $\gamma$ is required to achieve an economy of scale in one-way coordination. Lastly, the model predicts the lower limit of $\beta_{C_1}$ when $\gamma = 1$, signifying perfect learning, $\beta_{C_1}$ equals $\beta_{\text{mod}}$. Consequently, the maximum potential for economy of scale is determined by the modules within the interaction structure.

\section*{S5 Text: Mathematical model of coordination mechanism: Two-way coordination }

Here, we provide derivation of our mathematical model of coordination mechanism. First, number of modules for a project, $N_m$ scales with page size as:
\begin{equation} \label{eq:N_m}
    N_{\text{m}} \sim N^{\beta_{\text{mod}}}
\end{equation}
where \( 0 \leq \beta_{\text{mod}} \leq 1 \), is the scaling of number of modules with total contributors in project. $\beta_{mod} = 0$ suggests constant number of modules regardless of number of contributors in project, and $\beta_{mod} = 1$ suggests the number of modules linearly increases with number of contributors, leading to constant number of contributors in a module. Then, number of contributors in a module, $n$, is
\begin{equation} \label{eq:n}
    n = \frac{N}{N_{\text{m}}} \sim N^{1-\beta_{\text{mod}}}\;.
\end{equation}

Next, let number of potential interactions in module,
\begin{equation} \label{eq:Im}
    I_{\text{m}} \sim n^{\alpha}.
\end{equation}
with \( 1 \leq \alpha \leq 2 \), gives how the number of potential interactions scale with the module size. $\alpha = 2$ suggests well-mixed interaction within module. $\alpha = 1$ suggests each contributor interacts with a constant number of contributor regardless of module size. Then, total potential interactions for a page ($I$), is
\begin{align}
    I &\sim N_{\text{m}} I_{\text{m}} \\
    &= N^{\beta_{\text{mod}}}  N^{(1-\beta_{\text{mod}}) \alpha} \\
    &= N^{\beta_{\text{mod}} (1 - \alpha) + \alpha}.
\end{align}
Then, the exponent for total potential interaction, which is two-way coordination (reverts and talk page), is 
\begin{equation}\label{eq:c2}
    \beta_{C_2} = \beta_{mod} (1- \alpha) + \alpha.
\end{equation}

\section*{S6 Text: Mathematical model of coordination mechanism: One-way coordination }

We assume the editors learn through experience from past two-way coordination and learning accumulated. The more two-way coordination, the more editors learn. Given that we assumed interactions are bounded in a module, we consider learning are specific to a module as well. The amount of learning for a module increases with two-way coordination as, 
\begin{equation}
    k \sim (I_{\text{m}})^{\gamma}, 
\end{equation}
where $ 0 \leq \gamma \leq 1$, is the learning rate. $\delta = 0 $ suggests no learning, and $\gamma = 1$ suggests perfect learning---all experiences are learned. 

Then, we assume that the probability of escalating an interaction, $p_{e}$, is inversely proportional to learning, 

\begin{equation}
    p_{\text{e}}\propto 1/k.
\end{equation}
In other words, as learning doubles, the escalation probability half. Combining with Eqs. \ref{eq:n} and \ref{eq:Im}, we have 

\begin{align}\label{eq:P_e}
    p_{\text{e}} &\sim (I_{\text{m}})^{-\gamma}\\
     &\sim  N^{- \gamma(1-\beta_{mod})\alpha}.
\end{align}
Then, total number of escalations in all modules, $N_e$,  is the product of number of modules, interactions in each module, and the probability of escalation for each interaction: 
\begin{equation}
    N_{\text{e}} \sim N_{\text{m}} I_{\text{m}}  P_{\text{e}}.
\end{equation}
With Eq.\eqref{eq:N_m}, \eqref{eq:Im}, \eqref{eq:n}, \eqref{eq:P_e}, and simplify, we have
\begin{equation}
   N_e \sim N^{\beta_{mod}+ (1-\beta_{mod})\alpha(1-\gamma)}.
\end{equation}
Then, the exponent of the total escalation, which is one-way coordination, is 
\begin{equation}\label{eq:beta1}
    \beta_{C_1}= \beta_{mod} + (1-\beta_{mod}) \alpha (1-\gamma)
\end{equation}.

\section*{S7 Text: Mathematical model of coordination mechanism: Explaining residual }

Consider page $i$ with scaling residual of the two-way coordination, which is a measure of two-way coordination, denoted as $\xi_{C_2}^{(I)}$

Base the definition of scaling residual (\ref{eq:resi}), The number of interactions on page $i$ is: 
\begin{equation}
I^{(i)} \sim (N^{(i)})^{\beta_{C_2}} \exp(\xi^{(i)}_{C_2}),
\end{equation}
where $N^{(i)}$ is the number of contributors for page $i$. Combining Eq.\ref{eq:c2}, the number of interactions in each module is, 
\begin{equation}
I^{(i)}_m \sim (N^{(i)})^{(1-\beta_{mod})\alpha} \exp(\xi^{(i)}_{C_2}).
\end{equation}
Then, the escalation probability is 

\begin{align}\
    p_e^{(i)}&= 1/k^{(i)} \\
    &= (I^{(i)}_m)^{-\gamma}\\
    &= N^{-(1-\beta_{mod})\alpha\gamma}\exp(\xi^{(i)}_{C_2}). \\
\end{align}
Then, the number of escalated interaction, which is the amount of one-way coordination,  is
\begin{equation}
    N_e^{(i)} \sim N_m I_m^{(i)} Pe \sim (N^{(i)})^{\beta_1} \exp{\xi_{C_2}^{(i)} (1-\gamma)}
\end{equation}
Lastly, the scaling residual of one-way coordination is, 
\begin{equation}
    \xi_{C_1}^{(i)} \sim \log (Ne^{(i)}/(N^{(i)})^{\beta_1}) =\xi_{C_2}^{(i)} (1-\gamma)
\end{equation}
Since $0 \leq \gamma \leq 1$, the residual for admins and talk page should be positively correlated which is aligned with our observation in Tab.~\ref{table:resi}.
\clearpage

\begin{figure*}[!t]
    \centering
    \includegraphics[width=\textwidth]{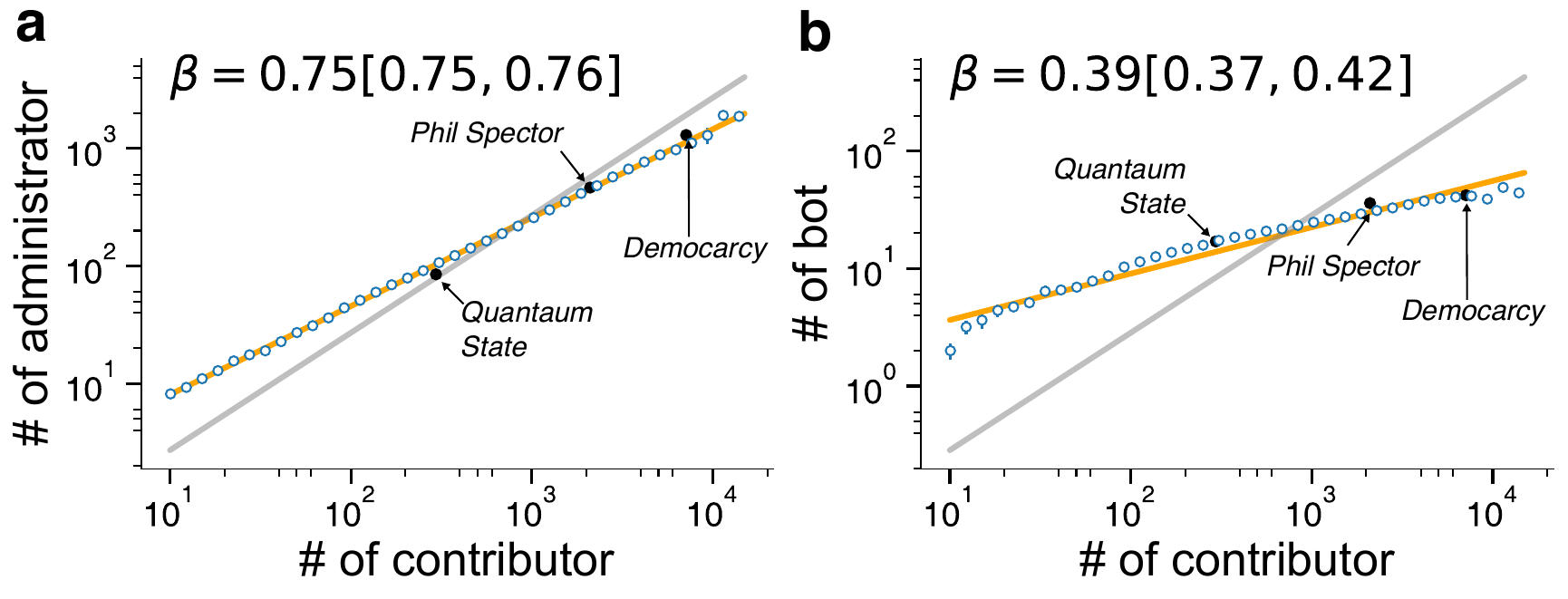}
    \caption{\textbf{Coordination costs in Wikipedia} The figures depict the scaling relationship of the required \textbf{(a)} number of administrators and \textbf{(b)} number of bots. The orange line represents the regression results, while the gray line indicates the baseline results with a scaling exponent of $\beta=1$. Blue dots represent the average value of each bin, with error bars denoting the standard error. }
    \label{fig:costs_number}
\end{figure*}

\section*{S8 Text:From weak interaction network to strong interaction network}

\section*{S9 Text: Why do we use cumulative totals for our unit of analysis?}

This integral approach, although seemingly unconventional in organization science, is a useful way to quantify the fundamental mechanisms at play because it not only effectively mitigates the impact of exogenous social events that may introduce idiosyncrasies depending on the chosen time-frame, but also enables the capture of hidden coordination costs that might be overlooked when examining isolated time slices of a page's history.

In the current paper, we choose an integral approach, moving beyond the direct, synchronous actions typically observed in organizational settings. 
Rather than merely counting the number of `active' contributors and their actions within a finite window of timeframe, typically a year or a month, we consider the system size as the cumulative total of all contributors who have ever made edits on a project page, as our baseline unit of analysis.
This unconventional methodology for assessing organization size offers unique insights and illuminates the perduring attributes of organizational structures over time. 
Our analysis will demonstrate that incorporating both the cumulative system size and related metrics does not alter the fundamental mechanisms at play. On the contrary, temporal ensembles (cumulative total) effectively diminish the influence of exogenous episodic perturbations by transient, singular social events, thereby uncovering the more fundamental characteristics of the system. 
Finally, the integral approach aims to reveal hidden coordination costs that might be overlooked in direct and synchronous interactions. This aspect is particularly crucial in the context of open-system and platform-based systems, where indirect interactions play a significant role.

\subsection*{Cumulative total as an ensemble of the dynamic systems}
The cumulative total is a way of estimating the ensemble average of the dynamic systems.
Let $t'$ be the segmented time frame, $N_t'$ denote the number of contributors at time $t'$, and $Y(N_t')$ represent the coordination cost (e.g., the number of reversions, length of the talk page) at time $t'$. In our framework, we examine the scaling relationship at time $t$ between $N_t = \int_0^t dt' Y(N_t')$ and $Y(N_t) = \int_0^t dt' Y(N_t')$, where $t$ is the observation point, yielding $Y(N_t) \sim N_t^{\beta_t}$. If we designate $\beta_{t'}$ as the scaling exponent $\beta$ for the segmented time $t'$ and $\beta_{t}$ as the scaling exponent $\beta$ of the integral approach, integrating both sides results in $\beta \simeq \beta_{t'}$. While this equality may vary for specific times due to temporal idiosyncrasies, the cumulative integration over time serves to mitigate such temporal variations. By accumulating interactions over time, we effectively eliminate temporal idiosyncrasies, thereby enhancing the robustness of our signals per page. This approach strengthens our ability to study the cumulative evolution of organization in Wikipedia, providing a more stable foundation for analyzing the intricate dynamics of coordination mechanisms.

\subsection*{The integral approach reveals hidden coordination costs}

The integral approach allows the capture of hidden coordination costs that would be overlooked when examining isolated time slices of a page’s history. Edits or coordination efforts made at specific time points can either raise or reduce coordination costs in the future. For example, contributors in the future might find that an earlier edit, once considered uncontroversial, has become controversial, leading to a surge in reverts and prolonged discussions on talk pages. Conversely, an earlier discussion might serve as a foundation for organizational learning, establishing a precedent applicable to future conflicts, thereby mitigating coordination costs over time. In essence, the rationale for employing the integral approach stems from the intricate nature of interactions among contributors across different time points in a page's entire history. Such temporal interdependence may obscure certain coordination costs, either incurred or inhibited by actions undertaken at different points in time, when viewed from a specific time window.

Furthermore, given the potential for previous contributors to re-engage in the production process during any future times, which is unlikely in conventional firms, it becomes even more crucial to consider the temporal interdependence among contributors within a page. Each contributor’s work of making edits to the current version of the page is not like simply adding a value to an already completed product left by previous contributors. Instead, it is more like an invitation for previous contributors to re-engage in the ongoing production process. Whenever new edits are made, previous contributors can potentially re-engage in the production process during any future periods to express their perspectives on the changes. To illustrate, while those who contributed in the Wikipedia page for “General Motors(GM)” in the past are always welcome to make contributions again on the same page, those who worked at GM in 1950s are typically not employed by GM in 2020s, which means there is no means they can join again the ongoing discussions in GM. This distinct nature of interactions in Wikipedia adds another reason to use the integral approach.

\begin{figure*}[t]
    \centering
    \includegraphics[width=0.7\textwidth]{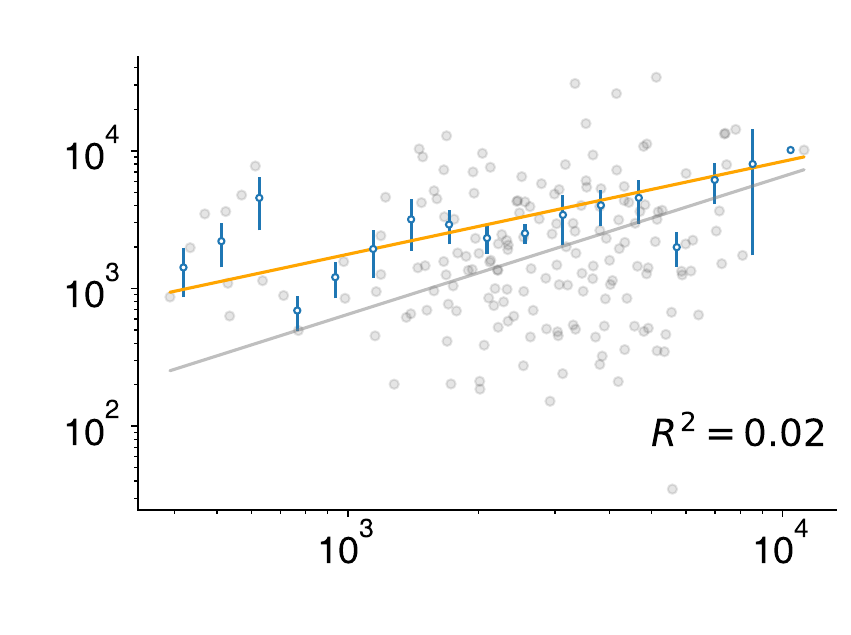}    \caption{\textbf{Coordination mechanisms in Wikipedia: FAQs} Within Wikipedia's talk pages, there exists a special section called FAQ (Frequently Asked Questions), which serves as a repository for past consensus, aimed at minimizing repetitive coordination efforts. Among the 48,822 vital pages in Wikipedia, a total of 175 pages have a FAQ section. Our analysis reveals that the length of these FAQ pages exhibits a sub-linear scaling relationship, indicating that it grows at a slower rate compared to the number of administrators (0.753).}
    \label{fig:faq}
\end{figure*}

\begin{figure*}[t]
    \centering
    \includegraphics[width=\textwidth]{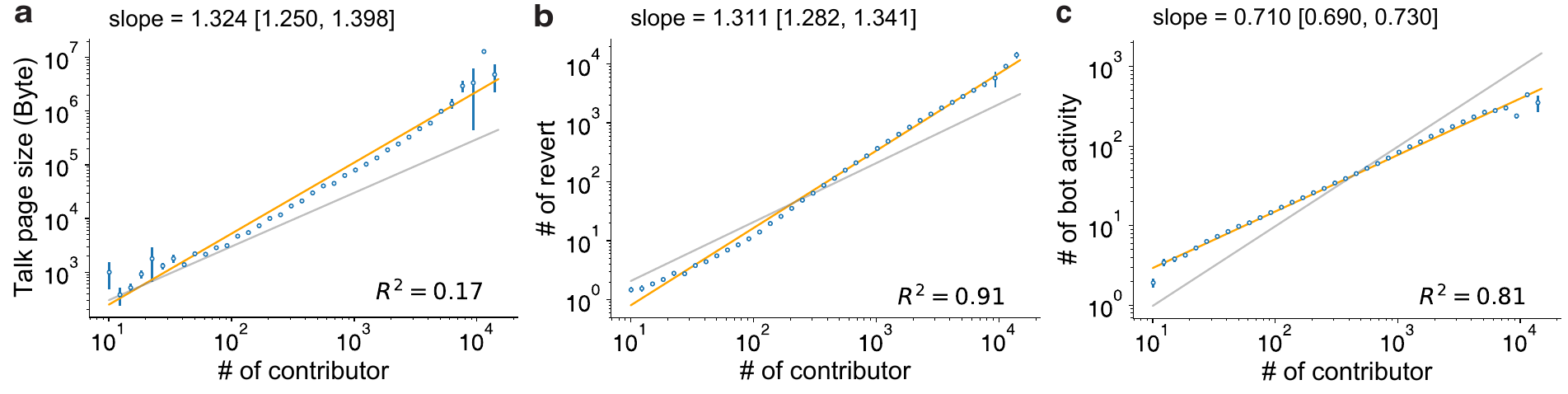}    \caption{\textbf{Robustness check of the scaling exponent} Out of the 48,822 vital pages in Wikipedia, 22,071 pages have not received any intervention from administrators. For robustness check, we examine the scaling relationship between the following metrics: \textbf{(a)} Talk page size, \textbf{(b)} number of reverts, and \textbf{(c)} number of bot activity and the number of contributors, including pages that do not have any edit from administrators. Our findings consistently demonstrate a super-scaling relationship on talk page size and the number of reverts, as well as a sub-linear scaling relationship on the number of bot activities. These results are in line with our previous observations.}
    \label{fig:costs_robust}
\end{figure*}

\begin{figure*}[!t]
    \centering
    \includegraphics[width=0.8\textwidth]{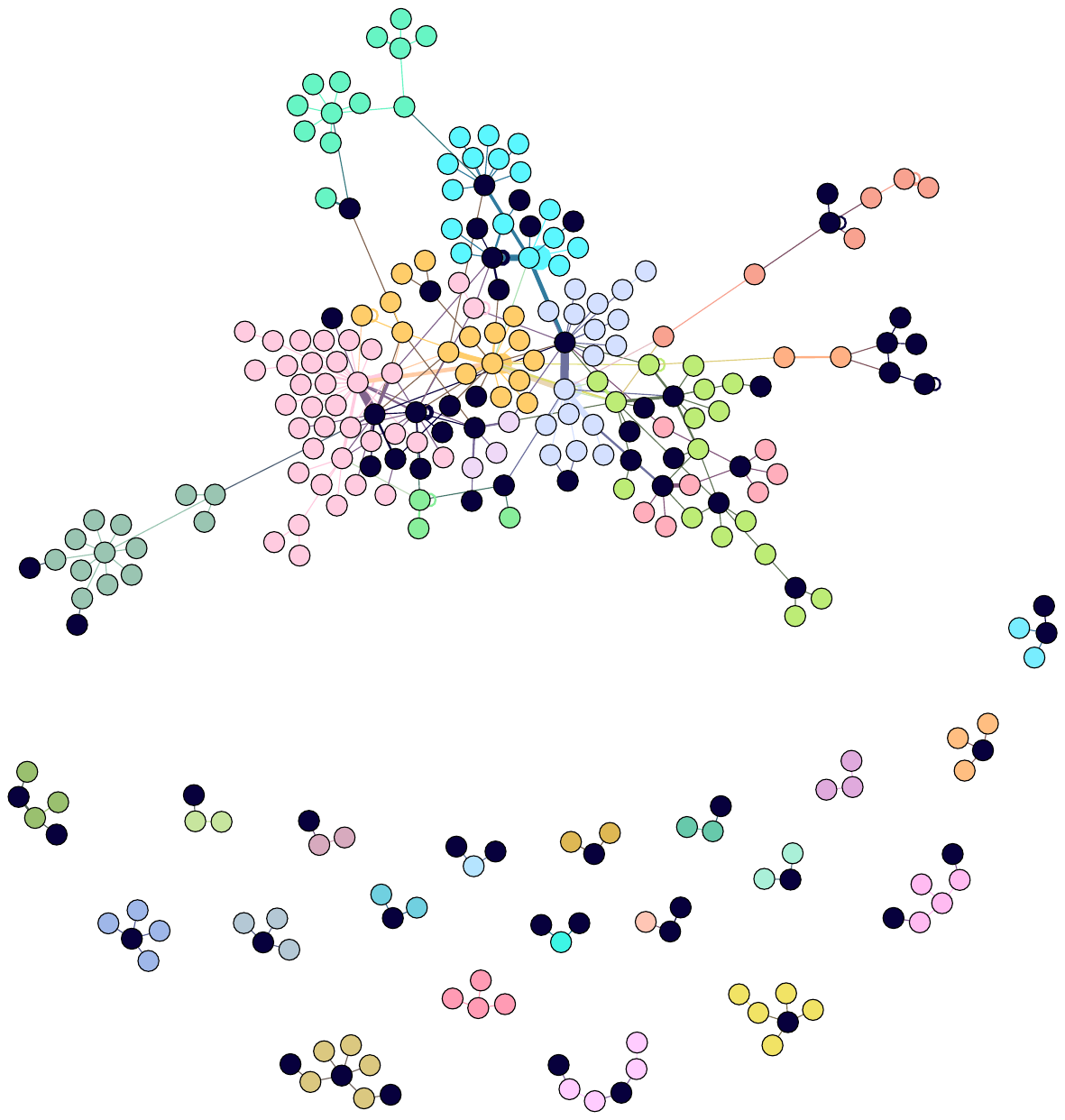}
    \caption{\textbf{Mention network found in the edit history of cold fusion} Color node indicates the extracted module with the community detection.  The red nodes are the admin and black nodes are bots.
     }
    \label{fig:coldfusion}
\end{figure*}

\begin{figure*}[!t]
    \centering
    \includegraphics[width=\textwidth]{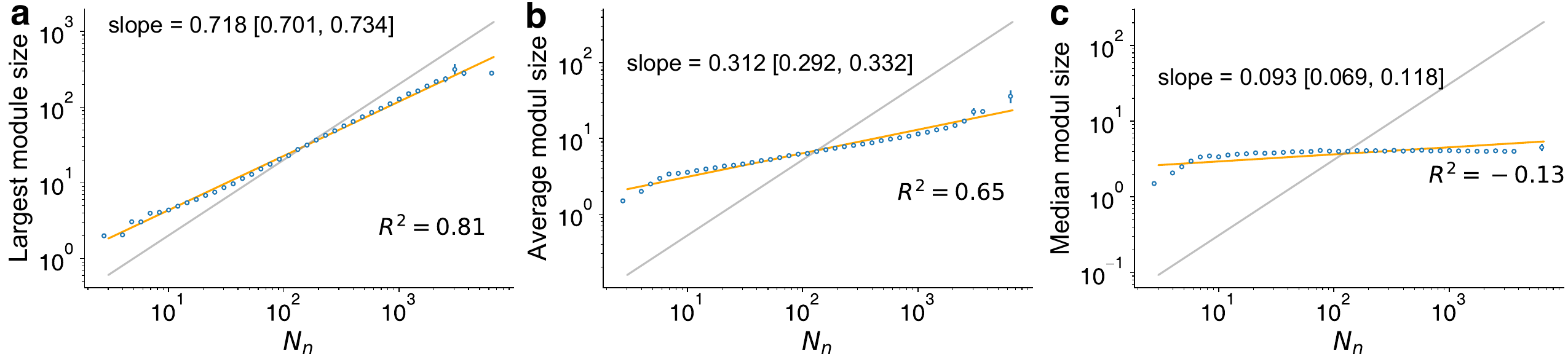}
    \caption{\textbf{Scaling relationship of the module size of mention network} We measure four measures for module size from the mention network: a) Maximum module size, b) Average module size c) Median module size. The orange line represents the regression results, while the gray line indicates linear scaling to help guide the eye. Blue dots represent the average value of each bin, with error bars denoting the standard error. 
     }
    \label{fig:mention}
\end{figure*}

\begin{figure*}[!t]
    \centering
    \includegraphics[width=0.8\textwidth]{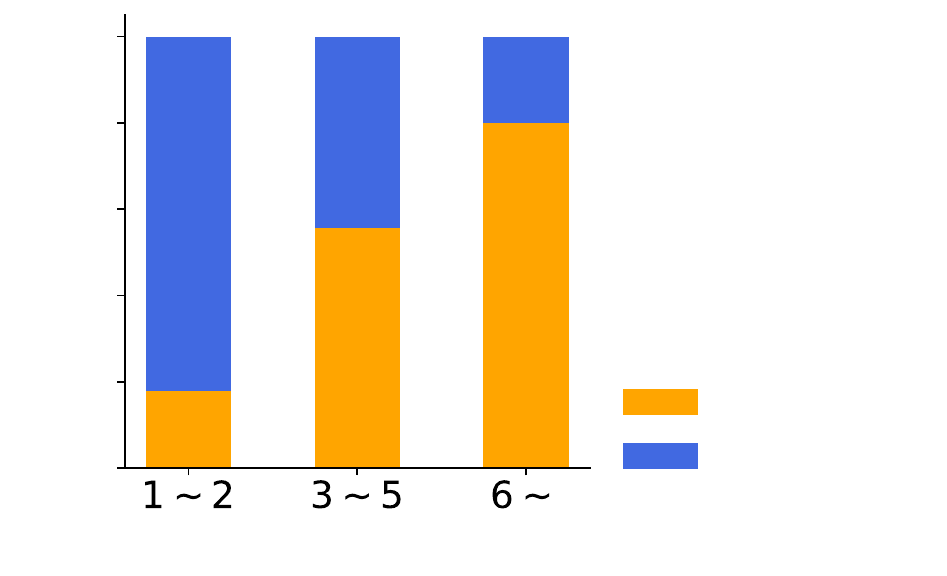}
    \caption{\textbf{Administrator as moderator} For each user, we calculate the unique number of adjacent modules, $\mathpzc{M}_i$. Administrators are more inclined to possess elevated $\mathpzc{M}_i$ values. This observation suggests a strong association between high $\mathpzc{M}_i$ values and the administrative role within the mention network. Furthermore, as the $\mathpzc{M}_i$ metric rises, the likelihood of a particular user holding the administrator role experiences a corresponding increase.
     }
    \label{fig:admin_role}
\end{figure*}




\begin{figure}[!t]
    \centering
    \includegraphics[width=0.8\textwidth]{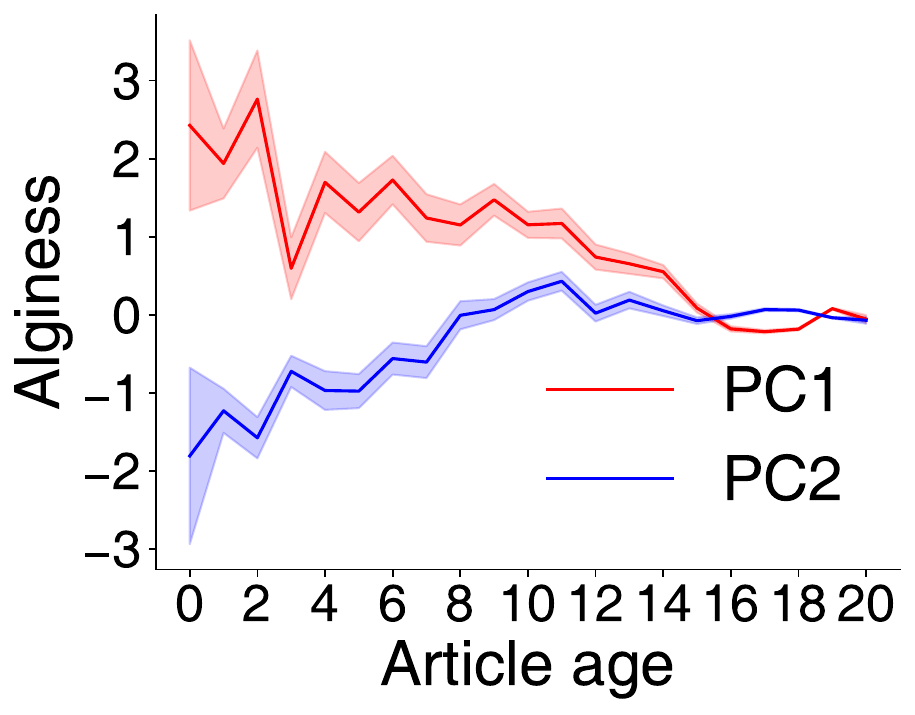}
    \caption{\textbf{Transition from personnel coordination to impersonal coordination}
   A strong alignment with PC2 (positive y) signifies a high level of impersonal coordination. Here, we use the article as the proxy of maturity. As the articles mature, their alignment with PC2 increases, indicating a seamless shift from personnel coordination to impersonal coordination. The standard errors are displayed below. We exclude COVID-related documents.
    }
    \label{fig:pc_scores}
\end{figure}

\begin{figure}[!t]
    \centering
    \includegraphics[width=\textwidth]{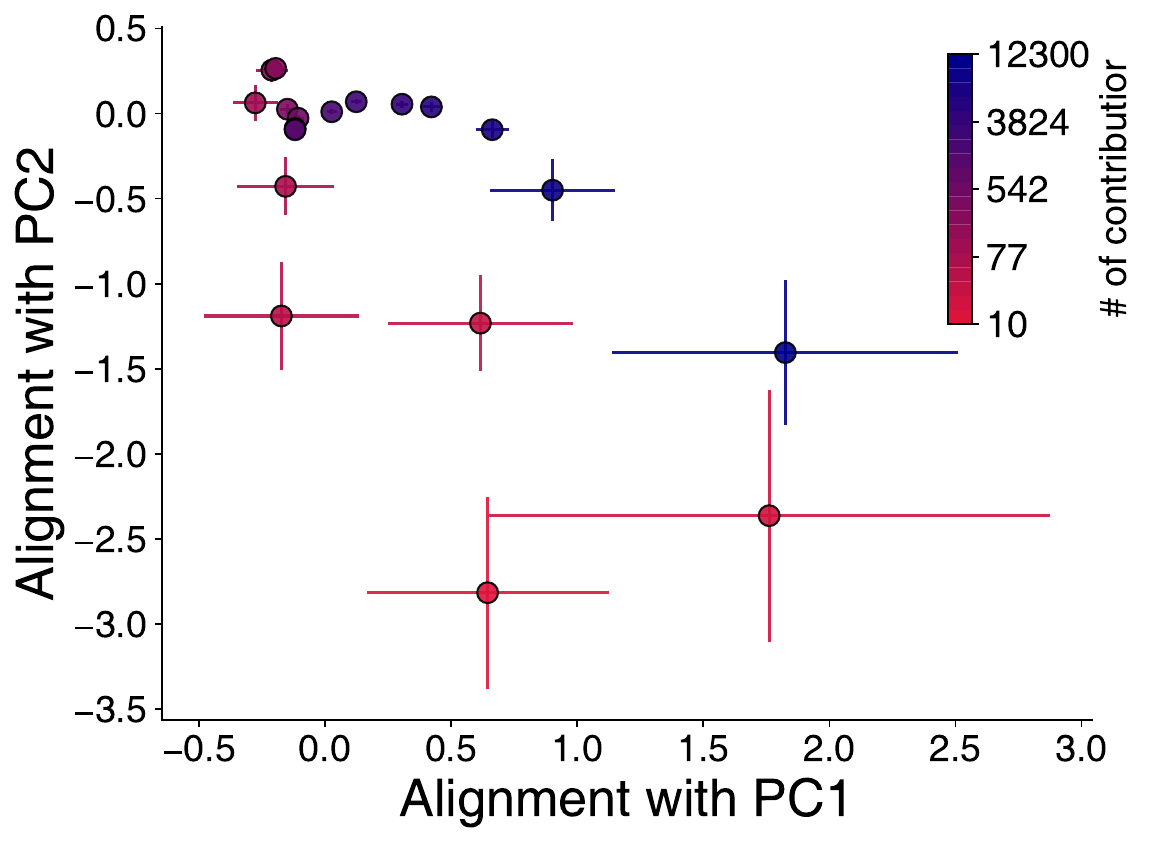}
    \caption{\textbf{Number of organization does not explain emergence of impersonal coordination } Dots and whiskers indicate the average value of each principal axis categorized by the binned number of contributors in the project.
    }
    \label{fig:real_size}
\end{figure}

\begin{sidewaystable}
\centering
\small
\begin{tabularx}{\textheight}{lrrrrrrrrr}
\toprule
      Category & Pages &$\beta^{\text{talk}}$ & $Y_0^{\text{talk}}$ &  $\beta^{\text{revert}}$ & $Y_0^{\text{revert}}$ &  $\beta^{\text{adm. act.}}$& $Y_0^{\text{adm. act.}}$ &  $\beta^{\text{bot act.}}$ &  $Y_0^{\text{bot act.}}$\\
\midrule
Arts             &   1,808 &  1.23 [1.08, 1.38] &  14.33 &  1.32 [1.23, 1.40] &  0.02 &  0.91 [0.86, 0.96] &  1.46 &  0.73 [0.70, 0.76] &  0.46 \\
Bio/Health       &   2,503 &  1.46 [1.29, 1.62] &   4.06 &  1.46 [1.41, 1.50] &  0.01 &  0.88 [0.83, 0.93] &  1.98 &  0.75 [0.71, 0.80] &  0.48 \\
Everyday life    &   1,553 &  1.12 [0.99, 1.25] &  31.70 &  1.31 [1.26, 1.37] &  0.02 &  0.92 [0.85, 0.99] &  1.24 &  0.72 [0.67, 0.77] &  0.47 \\
Geography        &   2,701 &  1.40 [1.27, 1.53] &   4.89 &  1.34 [1.27, 1.41] &  0.01 &  0.94 [0.90, 0.99] &  1.26 &  0.76 [0.73, 0.79] &  0.46 \\
History          &   1,913 &  1.53 [1.42, 1.63] &   4.68 &  1.24 [1.16, 1.32] &  0.03 &  0.90 [0.83, 0.97] &  1.96 &  0.76 [0.69, 0.82] &  0.48 \\
Math             &    455 &  1.10 [0.96, 1.24] &  79.99 &  1.36 [1.24, 1.48] &  0.01 &  0.92 [0.85, 0.99] &  1.33 &  0.80 [0.73, 0.87] &  0.25 \\
People           &   7,884 &  1.39 [1.31, 1.48] &   5.87 &  1.35 [1.29, 1.41] &  0.01 &  0.89 [0.83, 0.94] &  1.98 &  0.64 [0.62, 0.66] &  0.96 \\
Philos./Religion &    884 &  1.49 [1.38, 1.60] &   6.10 &  1.37 [1.33, 1.42] &  0.01 &  0.95 [0.88, 1.02] &  1.16 &  0.68 [0.63, 0.74] &  0.68 \\
Physical Sci.    &   1,936 &  1.44 [1.28, 1.61] &   6.06 &  1.30 [1.24, 1.37] &  0.02 &  0.96 [0.91, 1.02] &  1.09 &  0.76 [0.72, 0.79] &  0.47 \\
Society/Social   &   2,698 &  1.38 [1.27, 1.48] &   6.98 &  1.30 [1.24, 1.36] &  0.02 &  0.99 [0.95, 1.02] &  0.75 &  0.76 [0.72, 0.79] &  0.41 \\
Tech             &   1,679 &  1.21 [1.13, 1.30] &  17.54 &  1.29 [1.22, 1.36] &  0.02 &  0.97 [0.93, 1.02] &  0.84 &  0.81 [0.78, 0.85] &  0.26 \\
\midrule
Vital Article    &  26,014 &  1.33 [1.25, 1.41] &  11.57 &  1.28 [1.22, 1.33] &  0.02 &  0.91 [0.87, 0.95] &  1.71 &  0.69 [0.67, 0.71] &  0.66 \\
\bottomrule
  \end{tabularx}
\caption{\textbf{Scaling exponents of coordination costs by article category.} The scaling exponent $\beta$ and normalized constant $Y_0$ for all categories of Vital Articles, in all four measures of coordination costs. Talk page length and reverts correspond to mutual interactions, which is the two-way coordination. Administrator activity corresponds to supervision, and Bot activity corresponds to rule enforcement; both entail one-way coordination. Although there are minor differences in the scaling exponents across categories, the general trends --- superlinear  or two-way coordination or sublinear for one-way coordination --- remains consistent across all categories.}
\label{table:costs}
\end{sidewaystable}

\begin{table}[ht]
\begin{center}
\footnotesize
\begin{tabular}{lrrrr}
\toprule
{} &  $Y^{\text{talk}}$ &   $Y^{\text{revert}}$ &   $Y^{\text{adm act}}$ &   $Y^{\text{bot act}}$ \\
\midrule
$Y^{\text{talk}}$   &           &            &           &          \\
$Y^{\text{revert}}$ &          0.459 &             &           &       \\
$Y^{\text{adm act}}$  &          0.586 &            0.891 &           &          \\
$Y^{\text{bot act}}$   &          0.326 &            0.820 &           0.806 &         \\
\bottomrule
\end{tabular}
\caption{\textbf{Correlation between coordination costs} The Pearson correlation among the four coordination functions. The p-values for all correlations in this table are significant ($\ll0.001$).}
\label{table:raw}
\end{center}
\end{table}

\begin{table}[ht]
\begin{center}
\footnotesize
\begin{tabular}{lrrrr}
\toprule
{} &  $\xi^{\text{talk}}$ &   $\xi^{\text{revert}}$ &   $\xi^{\text{bot act.}}$ &    $\xi^{\text{adm. act.}}$ \\
\midrule
$\xi^{\text{talk}}$   &           &            &           &          \\
$\xi^{\text{revert}}$ &          0.092 &             &           &       \\
$\xi^{\text{bot act.}}$  &         0.080 &            0.088  &           &          \\
 \rowcolor{orange} $\xi^{\text{adm. act.}}$  &          0.386 &           0.229 &           0.427 &         \\
\bottomrule
\end{tabular}
\caption{\textbf{Correlation between residuals} The residuals of talk page size, reverts, bots' activities and admin activities are denoted as $\xi^{\text{talk}}$, $\xi^{\text{revert}}$, $\xi^{\text{bot act.}}$ and $\xi^{\text{adm. act.}}$, respectively.  Note that there are no correlations among residuals except for \textbf{admin activities}. The p-values for all correlations in this table are significant ($\ll0.001$).}
\label{table:resi}
\end{center}
\end{table}

\end{document}